\documentclass[epj,final]{svjour}

\usepackage{graphicx}
\usepackage{dcolumn}   
\usepackage{bm}        
\usepackage{amssymb}   
\usepackage{textcomp}
\usepackage{ucs}
\usepackage{hyperref}
\usepackage{amsmath}
\usepackage{color}
\usepackage{footnote}

\begin{document}

\title{Attenuation measurements of vacuum ultraviolet light in liquid argon revisited}

\author{A. Neumeier\inst{1} \and T. Dandl\inst{2} \and A. Himpsl\inst{2} \and M. Hofmann\inst{1,3} \and L. Oberauer\inst{1} \and W. Potzel\inst{1} \and S. Sch\"onert\inst{1} \and A. Ulrich\inst{2}\thanks{\emph{Andreas Ulrich:} andreas.ulrich@ph.tum.de 
}
}                     
\institute{Technische Universit\"at M\"unchen, Physik-Department E15, James-Franck-Str. 1, D-85748 Garching, Germany \and Technische Universit\"at M\"unchen, Physik-Department E12, James-Franck-Str. 1, D-85748 Garching, Germany \and now at: KETEK GmbH, Hofer Str. 3, 81737 M\"unchen, Germany}

\date{Published in NIM A (2015)}

\abstract{
The attenuation of vacuum ultraviolet light in liquid argon in the context of its application in large liquid noble gas detectors has been studied. Compared to a previous publication several technical issues concerning transmission measurements in general are addressed and several systematic effects were quantitatively measured. Wavelength-resolved transmission measurements have been performed from the vacuum ultraviolet to the near-infrared region. On the current level of sensitivity with a length of the optical path of 11.6\,cm, no xenon-related absorption effects could be observed, and pure liquid argon is fully transparent down to the short wavelength cut-off of the experimental setup at 118\,nm. A lower limit for the attenuation length of pure liquid argon for its own scintillation light has been estimated to be 1.10\,m based on a very conservative approach.
}           

\maketitle

\section{Introduction}\label{sec:introduction}
In a previous publication \cite{Neumeier_epjc_2012} we described measurements of the attenuation of vacuum ultraviolet (VUV) light in liquid argon in the context of particle detectors in rare event physics. A value for the attenuation length was derived, and a significant influence of xenon and other impurities was found and studied with good spectral resolution. Some experimental details, however, could not be addressed at that point. Meanwhile, the setup used in ref. \cite{Neumeier_epjc_2012}, has been moved to another laboratory in which also the emission of light from liquid rare gas samples is measured \cite{Heindl_epl,Heindl_jinst,neumeier_epl_ir,neumeier_epl_vuv_ir}. In addition, a gas system is used with an improved gas-purification technique as will be outlined below. In the following sections of this paper we will describe the systematic consideration of several issues which were only briefly discussed in ref.\,\cite{Neumeier_epjc_2012}.

\section{Experimental Details}

\begin{figure}
 \centering
 \includegraphics[width=\columnwidth]{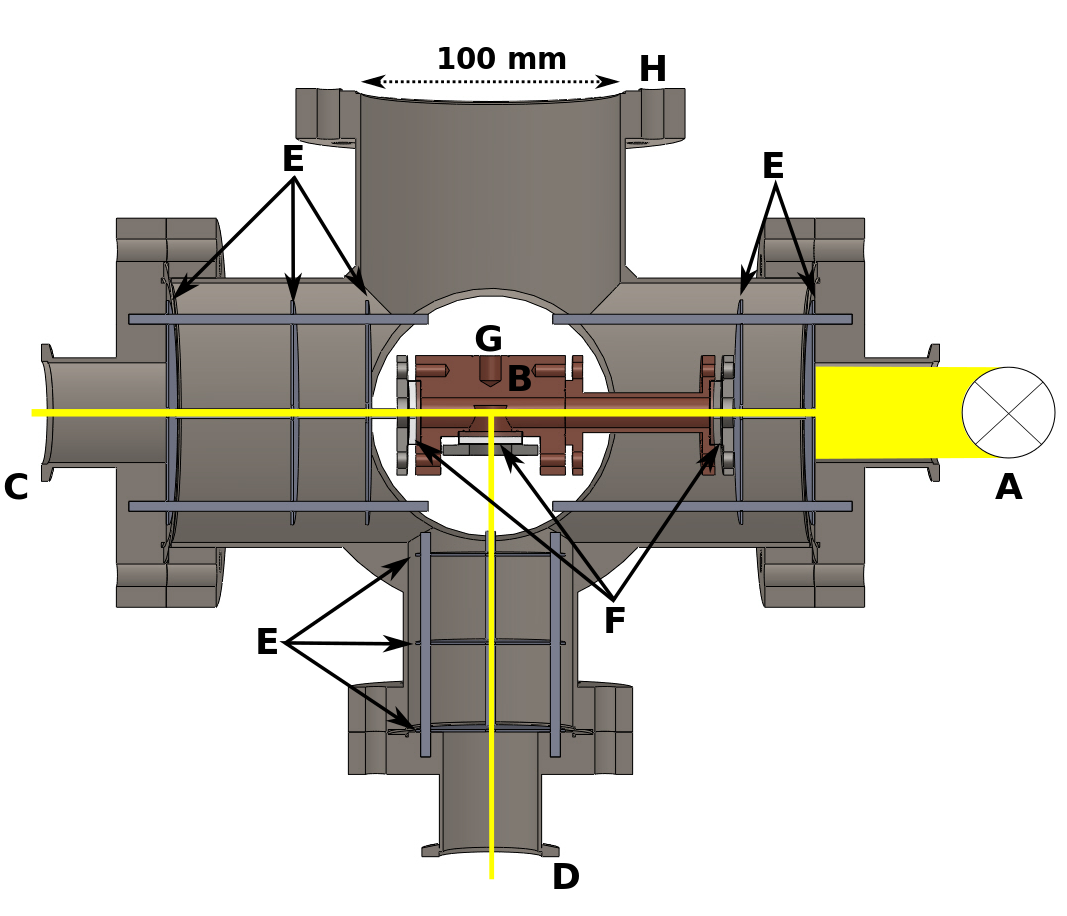}
 \caption{\textit{A cut through the experimental setup is shown. The transmission of liquid argon has been measured using a light source (A) which was attached to an evacuated CF-100 crosspiece (outer cell H). The light was sent through a cryogenic copper cell (inner cell or sample cell B) containing liquid argon. The transmitted light could be analyzed using a VUV spectrometer (McPherson 218 VUV monochromator with image intensified diode detector array) attached to flange C. The light entered and exited the sample cell through indium sealed MgF$_{2}$ windows (F). The optical path length the light had to traverse through the liquid was 11.6\,cm. Light scattered at an angle of 90\,\textdegree\, could be detected using a photodiode (Opto Diode Corp. model AXUV20A) attached to flange D at a distance of approximately 17\,cm and was read out by a nanoamperemeter. Straylight in the outer cell was avoided by apertures (E, diameter 4\,mm) in front of and behind the cryogenic sample cell, as well as in front of the photodiode. The sample cell was connected to a liquid nitrogen dewar via a connection G for cooling and positioning. For the measurements with a halogen lamp and a He-Ne laser, repectively, the flange (C) where the VUV monochromator was attached had been removed and a large diameter (8\,cm) regular optical glas window had been mounted instead.}}
 \label{fig:schnitt_zelle}
\end{figure}

 A cut through the experimental setup is shown in figure\,\ref{fig:schnitt_zelle}. The main modification of the inner cell containing the liquid argon sample compared to ref.\,\cite{Neumeier_epjc_2012} is an extension of its length from 5.8\,cm to 11.6\,cm for more sensitive attenuation measurements. The inner cell is mounted inside a vacuum cross piece of 100\,mm inner diameter (CF100). The cooling system was adapted from the experiment described in ref. \cite{Neumeier_epjc_2012}. A computer-controlled temperature regulation system was designed and built, which allows us now to regulate and stabilize the temperature with a precision of 0.1\,K within a narrow temperature range (from 84 to 87\,K for atmospheric pressure) to keep the argon in its liquid phase. It is based on a regulation loop (PI controller) where a platinum resistive temperature sensor (Pt\,100) provides the temperature information. The temperature regulation is performed by a 47\,$\Omega$, 20\,W resistor heating a rod (attached to G in figure\,\ref{fig:schnitt_zelle}), which holds the gas cell in place and is connected to a liquid nitrogen dewar.
 
\begin{figure}
 \centering
 \includegraphics[width=\columnwidth]{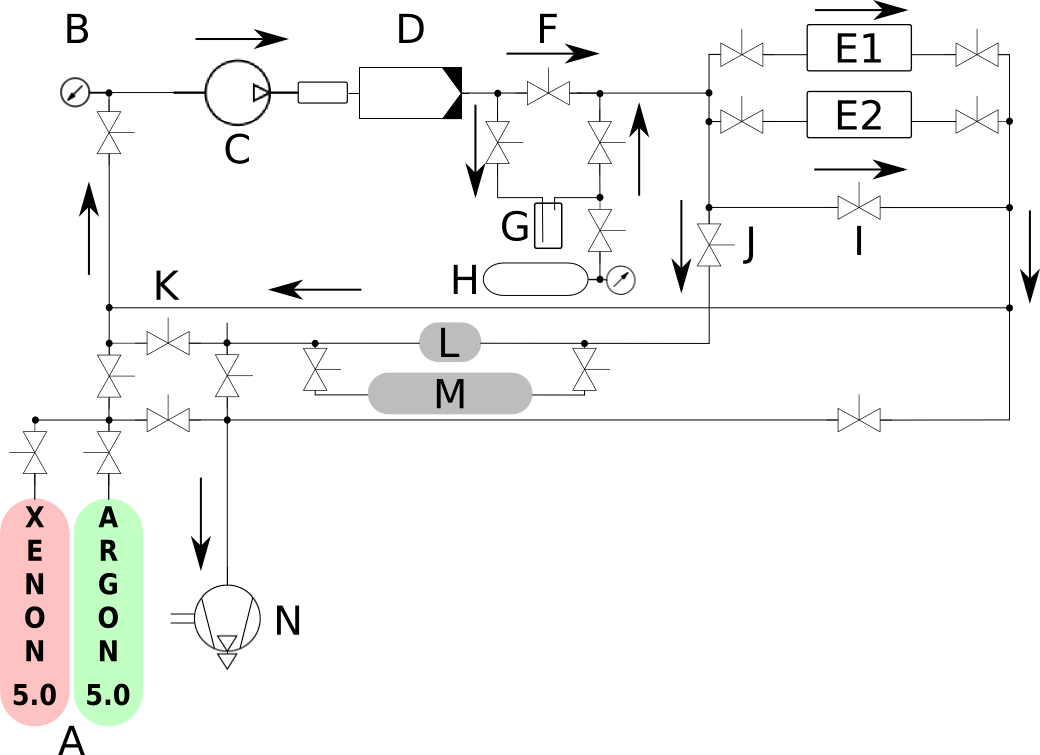}
 \caption{\textit{A schematic view of the gas system is presented. The gas (A) is filled into the system and its pressure is measured with a precise capacitive manometer (B; MKS Baratron 390H 1000). During the measurements the gas is continuously circulated by a metal bellows compressor (C) through a gas purifier (D; SAES getters, Model: MonoTorr Phase II, PS4-MT3-R2) and the experiments (E1, E2). In addition to the transmission measurements of liquid noble gas samples, this gas system is also used for studying the electron-beam induced scintillation of noble gases. Therefore, two experiments (E1 and E2) are indicated in the schematic view. The system also offers the possibility for a fractional distillation in a continuous flow mode by closing the bypass valve (F). The conically shaped distiller (G) can be immersed into liquid nitrogen thereby maintaining the distillation temperature by the depth of immersion. After the distillation, the distiller is closed and heated up again to enable the gas components in the distiller to expand into an expansion volume (H). The flow of gas through the experiment could be regulated by the bypass valves (I or J and K). Two gas reservoirs with different volumes (L: $\sim3l$, M:$\sim23l$) were used to adjust the amount of gas to measurements with different liquid volumes. The whole system could be evacuated by a turbo molecular pump (N).}}
 \label{fig:gassystem}
\end{figure}

\begin{figure}
 \centering
 \includegraphics[width=\columnwidth]{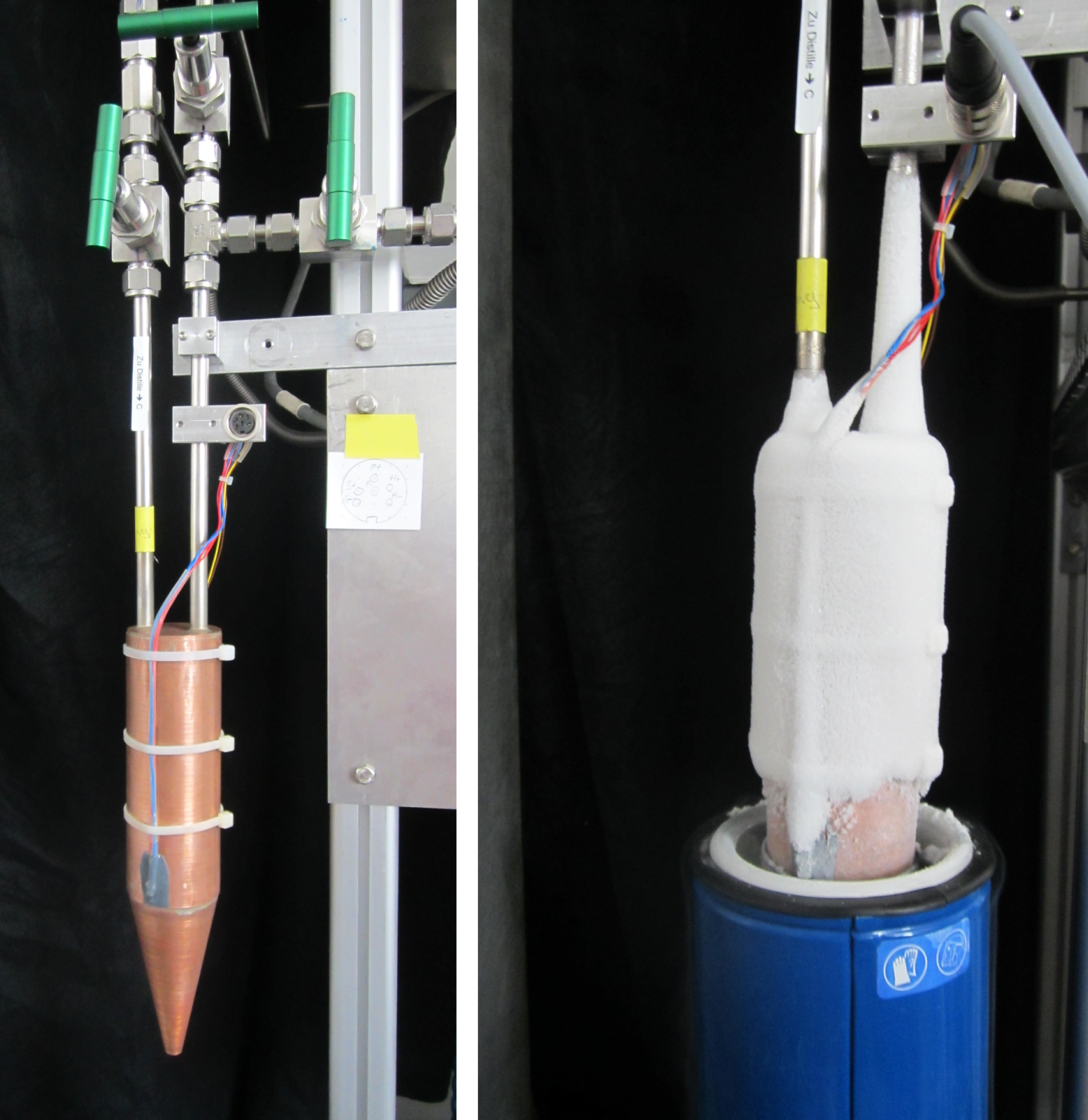}
 \caption{\textit{Left: The conically shaped distiller made of copper with gas inlet and gas outlet. A Pt-100 sensor is glued onto the distiller to monitor the temperature at the bottom of the distiller. Right: The distiller is immersed into a small dewar filled with liquid nitrogen and the gas is continuously circulated through the distiller. Note the different lengths of water frozen to the gas pipes indicating the flow direction from left to right.}}
 \label{fig:distille}
\end{figure}

The gas handling system is almost identical to the one described in ref. \cite{Heindl_jinst} and a schematic drawing can be seen in figure\,\ref{fig:gassystem}. Briefly, the gas is purified chemically by a commercial rare gas purifier (SAES model MonoTorr Phase II, PS4-MT3-R2) and circulated in a closed cycle with a metal bellow compressor. The gas is stored in a reservoir\footnote{Note that the reservoir is also part of the purification cycle.} from which it is condensed into the sample cell. For removing xenon, the gas sample is distilled before usage as in ref. \cite{Neumeier_epjc_2012}. In addition to the setup used in ref.\,\cite{Heindl_jinst}, a temperature controlled, conically shaped copper vessel could be immersed into liquid nitrogen to an appropriate extent for a careful temperature-controlled distillation process (see photograph in figure\,\ref{fig:distille}). The temperature of the distiller was stabilized $\sim0.5$\,K above the point of liquefication of argon.  This technique removes xenon impurities much better than the simple cylindrical container without gas flow which had been used before \cite{Neumeier_epjc_2012,neumeier_diploma_thesis}.

In the following the term "attenuation measurements" instead of "absorption measurements" is used since the effect of real absorption and light scattering in the sample can not be disentangled by just sending a light beam through the sample and comparing the exiting light for an empty and a filled sample cell, respectively. To test if there is a significant amount of stray- and scintillation light produced in the liquid argon under certain conditions, a VUV sensitive detector was mounted under an angle of 90\textdegree\ with respect to the beam of VUV light probing the sample. This VUV photodiode detector (Opto Diode Corp. model AXUV20A, attached to D in figure\,\ref{fig:schnitt_zelle}) could observe the sample from one side of the CF100 cross piece from a distance of approximately 17\,cm. It was read out by a current meter with sensitivity down to 20\,nA full scale. The VUV photodiode had a responsivity of 0.15\,$\frac{\text{A}}{\text{W}}$ \cite{XUV_Diode} at a wavelength of 127\,nm, thus, the optical sensitivity for reemitted scintillation light from liquid argon in a 90\textdegree angle was $\sim133$\,nW. However, in none of the measurements a significant signal has been measured by the VUV photodiode detector.
As in ref. \cite{Neumeier_epjc_2012} a deuterium arc lamp (Cathodeon Model V03) with MgF$_{2}$ window was used for most of the attenuation experiments described below. The VUV spectra with the sample cell evacuated and filled with liquid argon, respectively, were recorded with an image intensifier for VUV light mounted on a f=30\,cm vacuum monochromator (McPherson 218). For studying a geometrical effect which influences the measurements and which will be described in detail below, we extended the attenuation measurements in liquid argon also to the ultraviolet (UV), and in separate experiments to the visible (VIS) and near-infrared (NIR) region. The light sources in these measurements were a halogen lamp and, alternatively, also a 8\,mW He-Ne laser. The spectral response was measured using a small grating spectrometer with fiber optics coupling (Ocean Optics QE65000). A CCD camera (Atik 383Lc+) was used for measuring the beam profile, the beam position, and the integral intensity of the He-Ne laser beam for an empty and a filled sample cell. A large diameter (80\,mm) regular optical glass window was used at the port where the light exits the outer vacuum cell in the UV, VIS, and NIR experiments. With these modifications and extensions to the setup of refs. \cite{Neumeier_epjc_2012,neumeier_diploma_thesis} we have performed experiments to address various previously unanswered questions which had aroused during and after the previous work on light attenuation in liquid rare gases.

\section{Unphysical values for the Transmission ($>$\,1.0) in the Raw Data}

\begin{figure}
 \centering
 \includegraphics[width=\columnwidth]{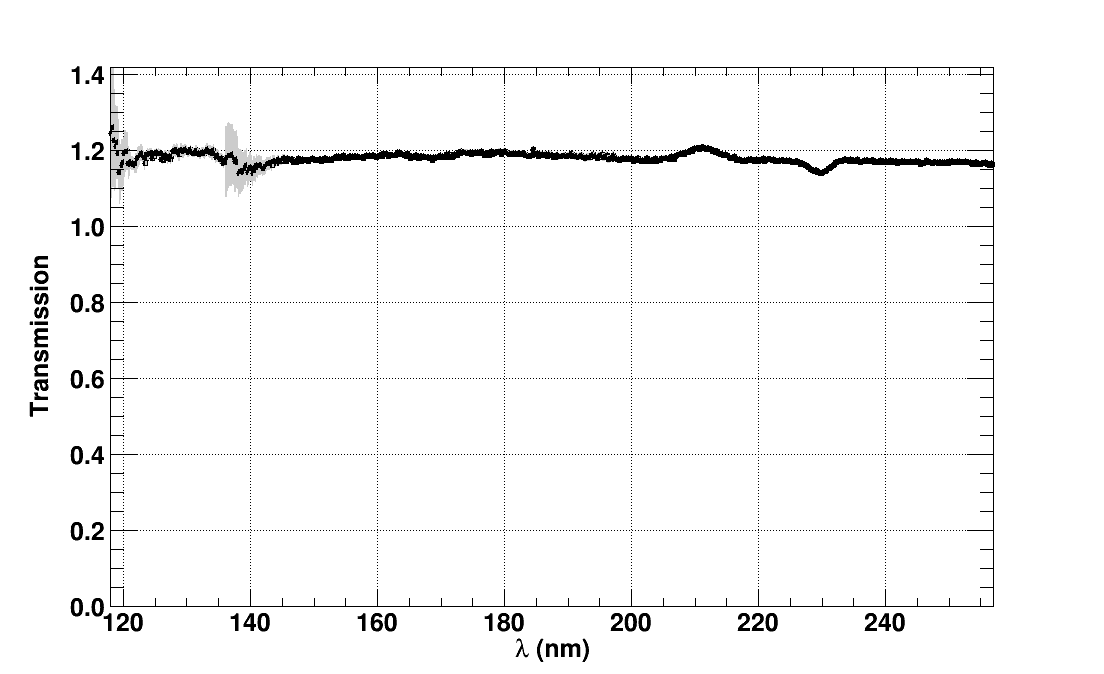}
 \caption{\textit{The measured transmission (black points with statistical (gray) error bars) of 11.6\,cm pure liquid argon is shown. Every analyzing step described in the text and in ref.\,\cite{Neumeier_epjc_2012} has been applied to the data except for a scaling to 1.0. With the current length of the inner cell of 11.6\,cm no absorption features were observed. The transmission is above a value of 1.0 due to the "Fresnel effect" and the finite divergence of the light source used (see discussion in subsections 3.1 and 3.2).}}
 \label{fig:transmission_lar}
\end{figure}

A spectrum showing the raw data of an attenuation measurement in liquid argon between 118 and 257\,nm is presented in figure\,\ref{fig:transmission_lar}. The measured transmission T is at non-physical values above 1.0 (T\,$\sim1.2$) over almost the whole wavelength range. Here, raw data means that two spectra obtained from the VUV image intensified diode array camera attached to the VUV monochromator were treated in the following way: The spectra were corrected for background, stray light, and the "fogging effect" (for details see ref.\,\cite{Neumeier_epjc_2012} and section \ref{sec:fogging_correction} in the present article), then divided and plotted without a scaling to 1.0. "Fogging effect" means attenuation due to a covering of the cold windows of the sample cell with traces of gas in the outer vacuum cell in which it is placed. This correction is reliable on a 5\% level which will be demonstrated in section \ref{sec:fogging_correction}. Two effects which can partly explain the increased transmission are described in the following subsections.

\begin{figure}
 \centering
 \includegraphics[width=\columnwidth]{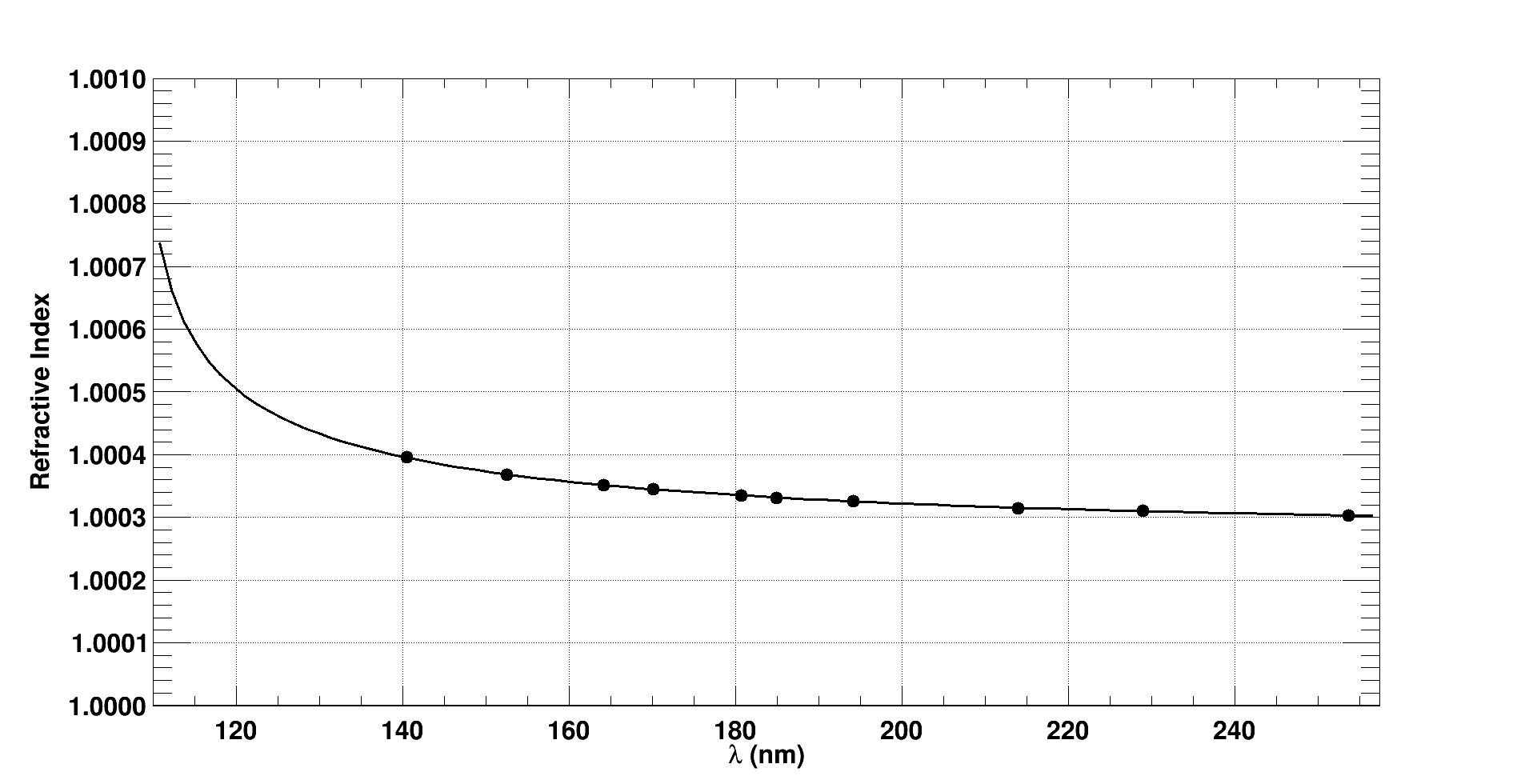}
 \caption{\textit{The wavelength-dependent refractive index of gaseous argon (273K, 1013\,mbar) adapted from Bideau-Mehu et al. \cite{Bideau-Mehu} is shown. The black dots indicate measured values and the black line shows a model calculation also adapted from ref.\,\cite{Bideau-Mehu}.}}
 \label{fig:Brechungsindex_GAr}
\end{figure}

\begin{figure}
 \centering
 \includegraphics[width=\columnwidth]{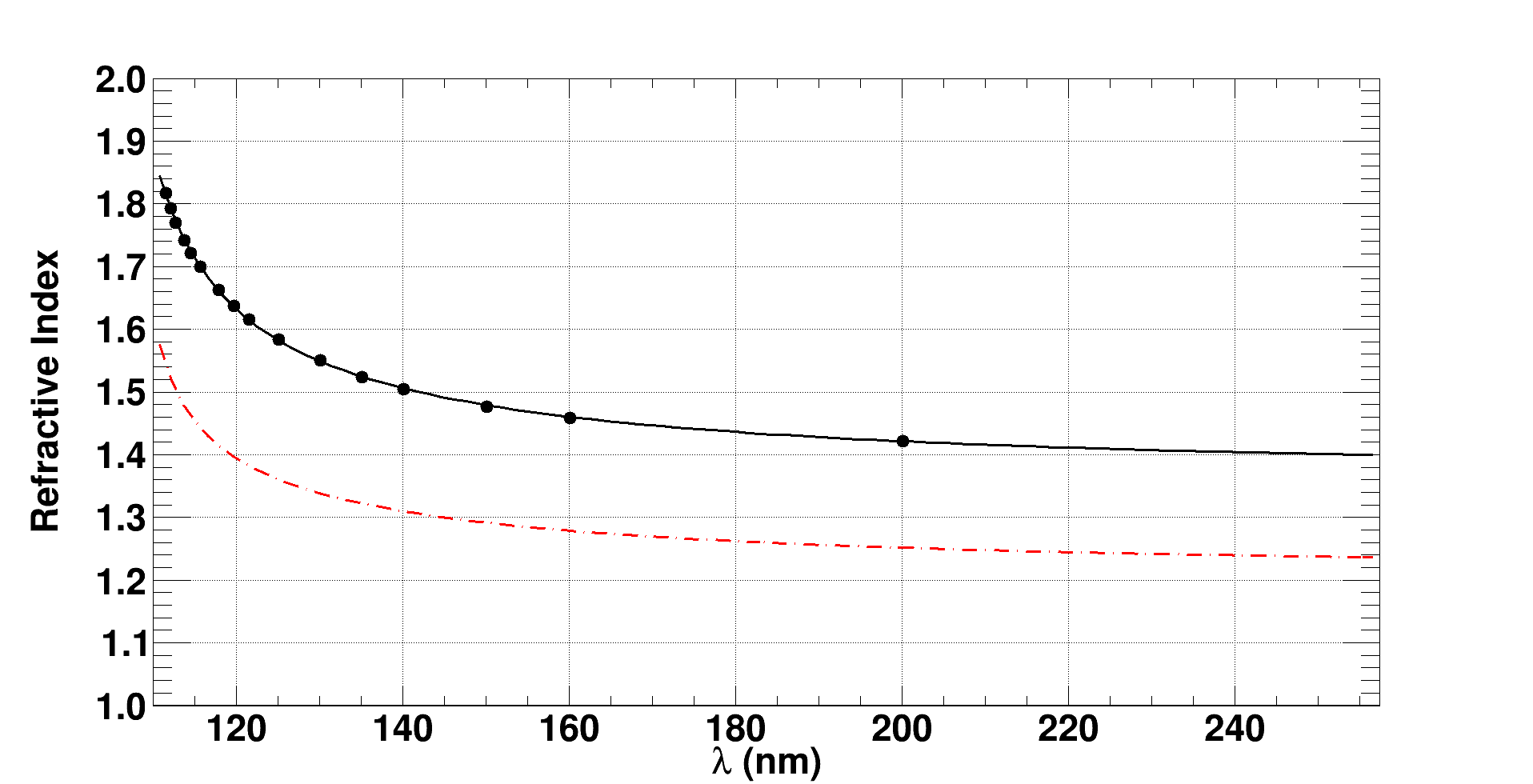}
 \caption{\textit{The wavelength-dependent refractive index of MgF$_{2}$ at 80\,K (black curve) adapted from Laporte et al.\,\cite{laporte} is shown. The black dots show measurements and the black line shows a model fit \cite{laporte} to the data. The red (colour online) dashed-dotted curve shows the wavelength-dependent refractive index of liquid argon obtained from a density scaling of the data in figure\,5 in analogue to the strategy presented in ref.\,\cite{Antonello}.}}
 \label{fig:Brechungsindex_LAr_MgF2}
\end{figure}

\paragraph{3.1 Fresnel Effect}
\label{subsec:fresneleffect}
The attenuation spectrum in figure\,\ref{fig:transmission_lar} shows that the transmission data is above 1.0 over the whole wavelength range from 118\,nm to 257\,nm. In analogy to the gas phase\footnote{Transmission measurements in gaseous argon ($\sim1000$\,mbar and room temperature) showed a transmission of 1.0 for the whole range of the experimental setup (118 - 257\,nm).}, purified liquid argon is assumed to be highly transparent at long wavelengths. Therefore, we had scaled our transmission spectra to a transmission value of 1.0 for wavelengths longer than 200\,nm in ref. \cite{Neumeier_epjc_2012}. An obvious effect which justifies such a correction is the modified reflection of light at the two inner surfaces of the MgF$_{2}$ windows of the sample cell when the cell is filled with liquid argon with an index of refraction different from 1.0 ("Fresnel formulas"\footnote{Due to the higher refractive index of liquid argon compared to vacuum used in the reference measurement, less light is reflected. Hence, more light is transmitted at the inner surfaces of the MgF$_{2}$ windows, when the sample cell is filled with liquid argon.}). Figure\,\ref{fig:Brechungsindex_GAr} shows the wavelength-dependent refractive index of gaseous argon at standard conditions (273\,K, 1013\,mbar) measured \cite{Bideau-Mehu} at ten different wavelengths and a model curve describing the wavelength dependence also adapted from \cite{Bideau-Mehu}. A wavelength-dependent value for the index of refraction of liquid argon has been published in ref.\,\cite{Antonello} based on the assumption that values available for the gas (see figure\,\ref{fig:Brechungsindex_GAr} and ref.\,\cite{Bideau-Mehu}) can be converted into values for the liquid just taking the density change into account. The wavelength-dependent refractive index of MgF$_{2}$ at cryogenic temperatures has been adapted from ref.\,\cite{laporte}. For comparison Figure\,\ref{fig:Brechungsindex_LAr_MgF2} shows the wavelength resolved refractive index of MgF$_{2}$ and liquid argon. 

\begin{figure}
 \centering
 \includegraphics[width=\columnwidth]{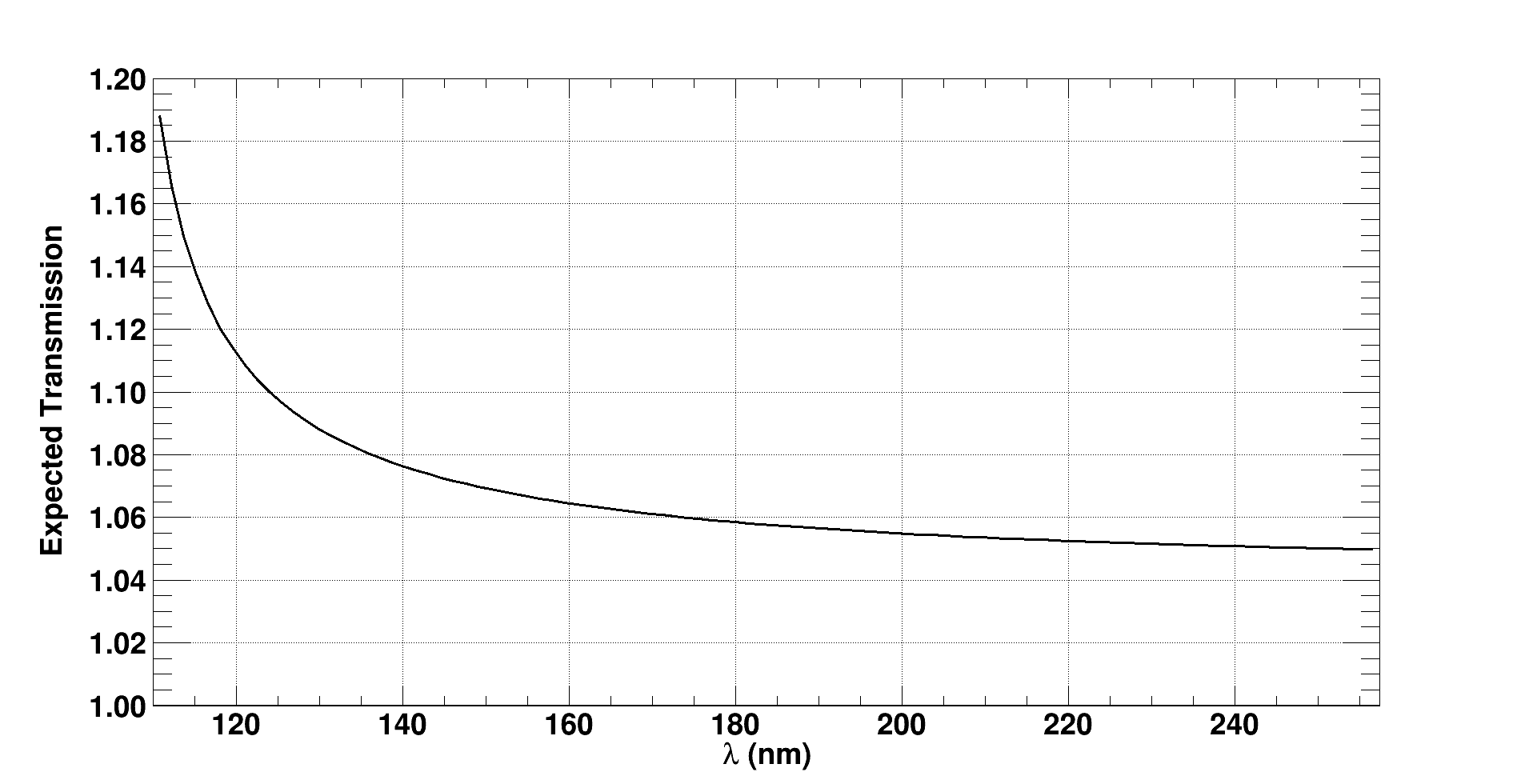}
 \caption{\textit{A calculation of the expected transmission ("Fresnel Effect") of the inner cell filled with liquid argon without any absorption or attenuation features is presented. This calculation has been performed using eq.\,1 and only takes into account the reflection for normal incidence of light at the two inner surfaces of the MgF$_{2}$ windows of the sample cell. The data for the wavelength-dependent index of refraction for MgF$_{2}$ and liquid argon are adapted from refs.\,\cite{Antonello} and \cite{laporte}, respectively.}}
 \label{fig:fresnel_senkrecht}
\end{figure}

The expected transmission spectrum from the "Fresnel effect" alone (without any absorption or attenuation features) for normal incidence of light through the cryogenic MgF$_{2}$ windows is shown in figure\,\ref{fig:fresnel_senkrecht} and calculated using equation\,\ref{eq:fresnel_senkrecht}.

\begin{equation}
\label{eq:fresnel_senkrecht}
 \text{T}_{\text{Fr}} = \left(\frac{1-\left(\frac{\text{n}_{\text{MgF}_{2}}-\text{n}_{\text{LAr}}}{\text{n}_{\text{MgF}_{2}}+\text{n}_{\text{LAr}}}\right)^{2}}{1-\left(\frac{\text{n}_{\text{Vak}}-\text{n}_{\text{MgF}_{2}}}{\text{n}_{\text{Vak}}+\text{n}_{\text{MgF}_{2}}}\right)^{2}}\right)^{2}
\end{equation}

\begin{description}
  \item [$\text{T}_{\text{Fr}}=\text{T}_{\text{Fr}}(\lambda)$] Wavelength-dependent transmission without any absorption or attenuation features
  \item [$\text{n}_{\text{MgF}_{2}}=\text{n}_{\text{MgF}_{2}}(\lambda)$ ] Wavelength-dependent refractive index of magnesium fluoride
  \item [$\text{n}_{\text{LAr}}=\text{n}_{\text{LAr}}(\lambda)$ ] Wavelength-dependent refractive index of liquid argon
  \item [$\text{n}_{\text{Vak}}=1$ ] Refractive index of vacuum
\end{description}

However, assuming that the refractive index of liquid argon is realistic, we find that the "Fresnel correction" for normal incidence of light can not describe the transmission data shown in figure\,\ref{fig:transmission_lar} quantitatively\footnote{A comparison of figure \ref{fig:transmission_lar} and figure \ref{fig:fresnel_senkrecht} at a wavelength of e.g. 180\,nm shows that only 6\,\% instead of 20\,\% transmission above a value of 1.0 can be explained by the "Fresnel correction".}. This indicates that at least one further process contributes, induced by the finite divergence of the light source, which is described in the next subsection.

\paragraph{3.2 Finite Divergence of the Light emitted from the Light Source}
\label{subsec:divergenceeffect}

\begin{figure}
 \centering
 \includegraphics[width=\columnwidth]{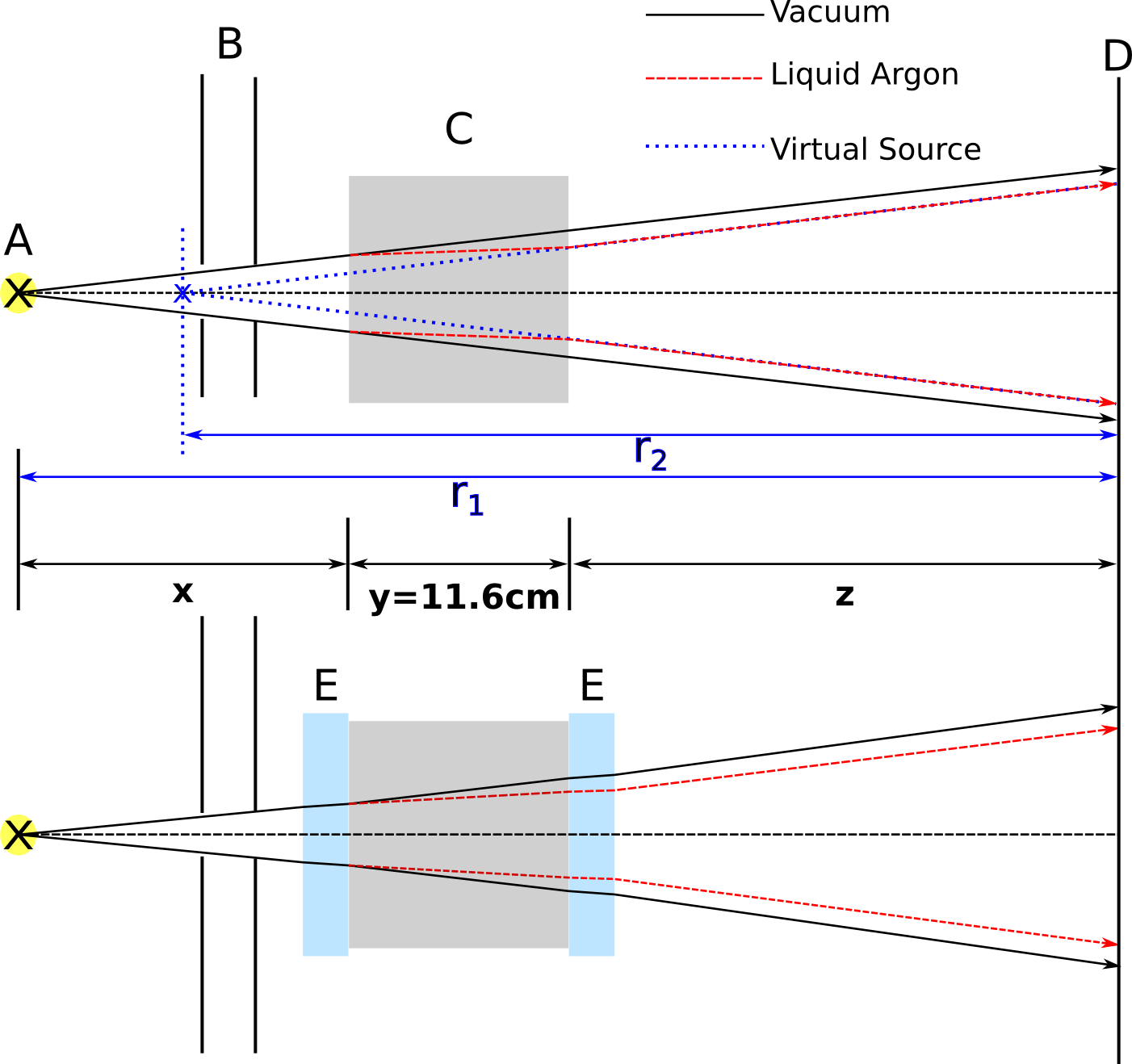}
 \caption{\textit{The upper scheme shows a simplified schematic drawing of the experimental setup without MgF$_{2}$ windows to explain the influence of a changing medium (C) on the beam profile. A light source (A) with finite divergence is placed at a distance $r_{1}$ from the detector plane (D). The apertures (B) had a diameter of 4\,mm. The black arrows indicate the beam profile for the case when the inner cell is empty (evacuated). The red (colour online, dashed) arrows indicate the beam profile when the inner cell is filled with a medium with a refractive index larger than 1.0 (e.g., liquid argon). The narrowing of the beam profile leads to a higher intensity at the detector plane (D). This can be treated in analogy to a virtual source which has been moved from an initial distance $r_{1}$ to a new distance $r_{2}$ relative to the detector plane. The intensity at the detector plane is therefore enhanced by a factor of $\left(\frac{r_{1}}{r_{2}}\right)^{2}$. The lower drawing shows a more detailed schematic where the MgF$_{2}$ windows (E, thickness 5\,mm) have been taken into account. In the case of the deuterium source the distances were: x\,=\,17.5\,cm, y\,=\,11.6\,cm and z\,=\,36.5\,cm.}}
 \label{fig:fokussierung_schematisch}
\end{figure}

A second effect which occurs when the probing light has a finite divergence has to be considered to explain the excessive transmission values. This geometrical effect (visualized by the schematic drawing in figure\,\ref{fig:fokussierung_schematisch}) acts as if filling the sample cell leads to a displacement of the light source closer to the detector behind the cell (at a distance z), leading to an increase in intensity. The impact of this effect can become very significant when light of finite divergence is used to probe the attenuation in a long sample cell. If a point-like light source is placed at a distance x from the entrance window of an absorption cell, the light travels more collimated through the cell when it is filled with a medium with an index of refraction $\text{n}>1$ for the full length of the cell y. Exiting the cell (filled as well as empty) the light propagates with the same divergence as in front of the cell. However, it has to be mentioned that the integral amount of transmitted light is not affected by the "finite divergence effect". This effect only becomes visible when restricting optical elements (e.g. apertures) are positioned in the optical path between sample cell and detector, or the active area of the detector is smaller than the area of the beam profile of the transmitted light (e.g. monochromator slits). In the present experiment both is the case which artificially increase the measured transmission. Using the geometrical effect according to the upper drawing of figure\,\ref{fig:fokussierung_schematisch}, formula \ref{eq:strahlfokussierung} has been derived: 

\begin{equation}
 \frac{I_{1}}{I_{2}} \sim \left(\frac{r_{1}}{r_{2}}\right)^{2}=\left(\frac{x+y+z}{x+\frac{y}{n_{\text{LAr}}}+z}\right)^{2}
 \label{eq:strahlfokussierung}
\end{equation}

\begin{figure}
 \centering
 \includegraphics[width=\columnwidth]{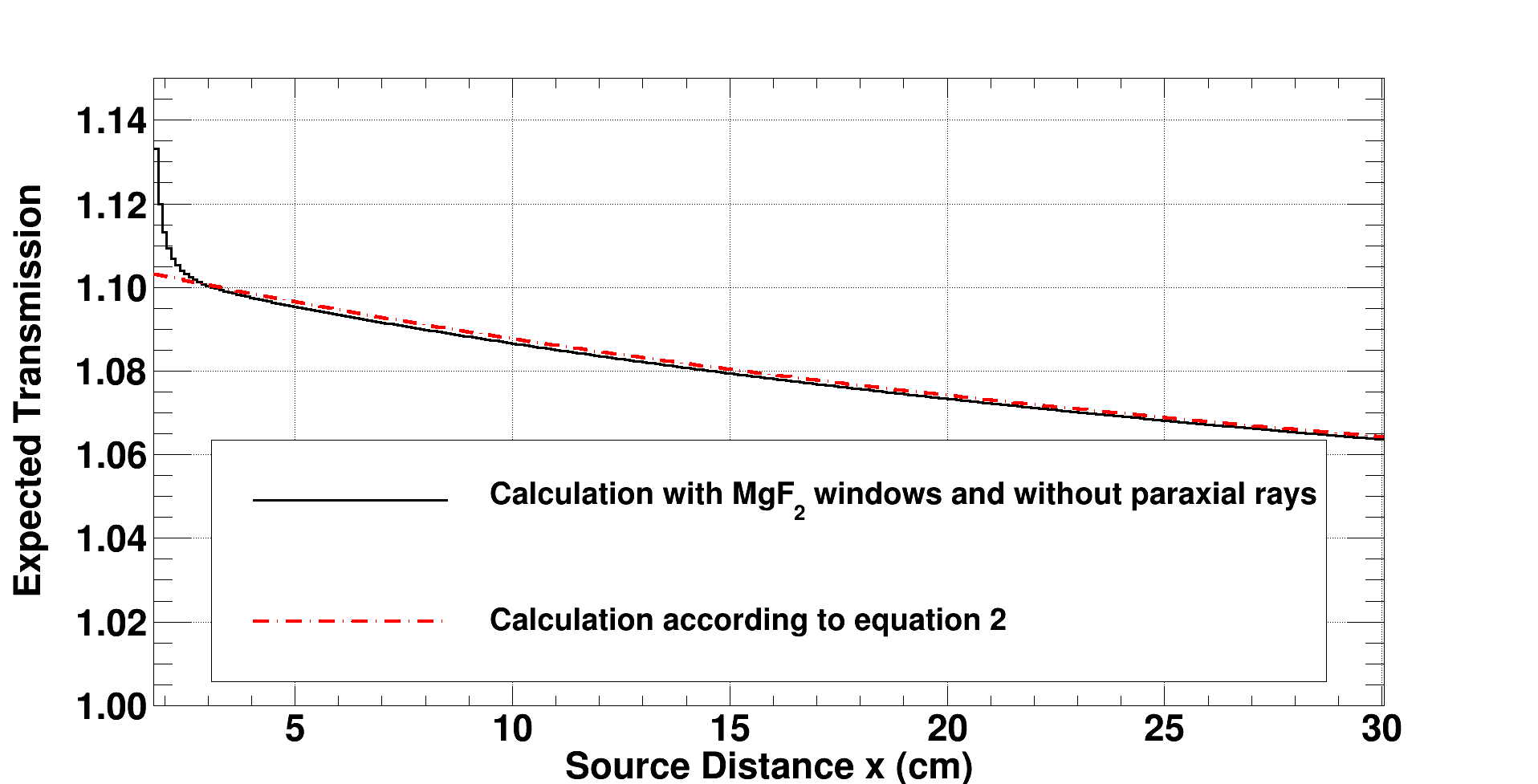}
 \caption{\textit{The expected transmission ("finite divergence effect") of liquid argon at 180\,nm without any attenuation, absorption or Fresnel effects is calculated. A light source with a finite divergence leads to an enhanced transmission. The red (colour online) dashed line shows the excessive transmission values which have been calculated using the approximate (paraxial rays, no MgF$_{2}$ windows) eq.\,2 for different source distances x (see figure\,8 upper drawing). The black solid line shows a detailed calculation (in one millimeter steps) of the same effect without paraxial rays and taking also into account the MgF$_{2}$ windows (see figure\,8 lower drawing). The deuterium lamp in the experimental setup had a distance x of 17.5\,cm. Consequently, the approximation in the upper drawing of figure\,8 together with equation\,2 nicely describe the effect for the geometry used in this experiment.}}
 \label{fig:fokussierungseffekt}
\end{figure}

The position of the virtual source and the $\left(\frac{r_{1}}{r_{2}}\right)^{2}$ law for the intensity ratio were used to quantify the the intensity ratio $\left(\frac{I_{1}}{I_{2}}\right)$; $\text{r}_{1}$ and $\text{r}_{2}$ are the distances from the detector to the real and virtual source, respectively. It could be shown that the finite thickness of the optical windows of 5\,mm could be neglected in comparison with the experimental errors (see schematic figure\,\ref{fig:fokussierung_schematisch} and the results of the calculations in figure\,\ref{fig:fokussierungseffekt}). The approximation for paraxial rays ($\text{sin}(\alpha)\sim\alpha$) shows that the effect does not disappear for that case and is actually independent of the angles. Positioning the deuterium light-source at a larger distance from the sample cell would result in a reduced transmission due to a reduced "finite divergence effect". Increasing the distance between deuterium light-source and detector also leads to a decreasing signal at the detector. 

A calculation\footnote{Note that this calculation has been performed with the refractive index of liquid argon at a wavelength of 180\,nm. It can be expected that the wavelength-dependent refractive index of liquid argon which has been derived from ref.\,\cite{Antonello} is more accurate at longer wavelengths than at short wavelengths (closer to the resonance lines) since these data are based on a density scaling of the refractive indices from the gas phase into the liquid phase. At 180\,nm, a measured value of the refractive index of gaseous argon \cite{Bideau-Mehu} is available. Below 140\,nm no measured values of the refractive index of gaseous argon are available (see figure \ref{fig:Brechungsindex_GAr} and ref.\,\cite{Bideau-Mehu}), which leads to a situation where a density scaling from the gas into the liquid phase is performed without any measurements in the gas phase.} of the expected transmission due to the finite divergence of the light source without the Fresnel effect (see subsection 3.1) and without absorption or attenuation effects for different source distances x is shown in figure\,\ref{fig:fokussierungseffekt}.

\begin{figure}
 \centering
 \includegraphics[width=\columnwidth]{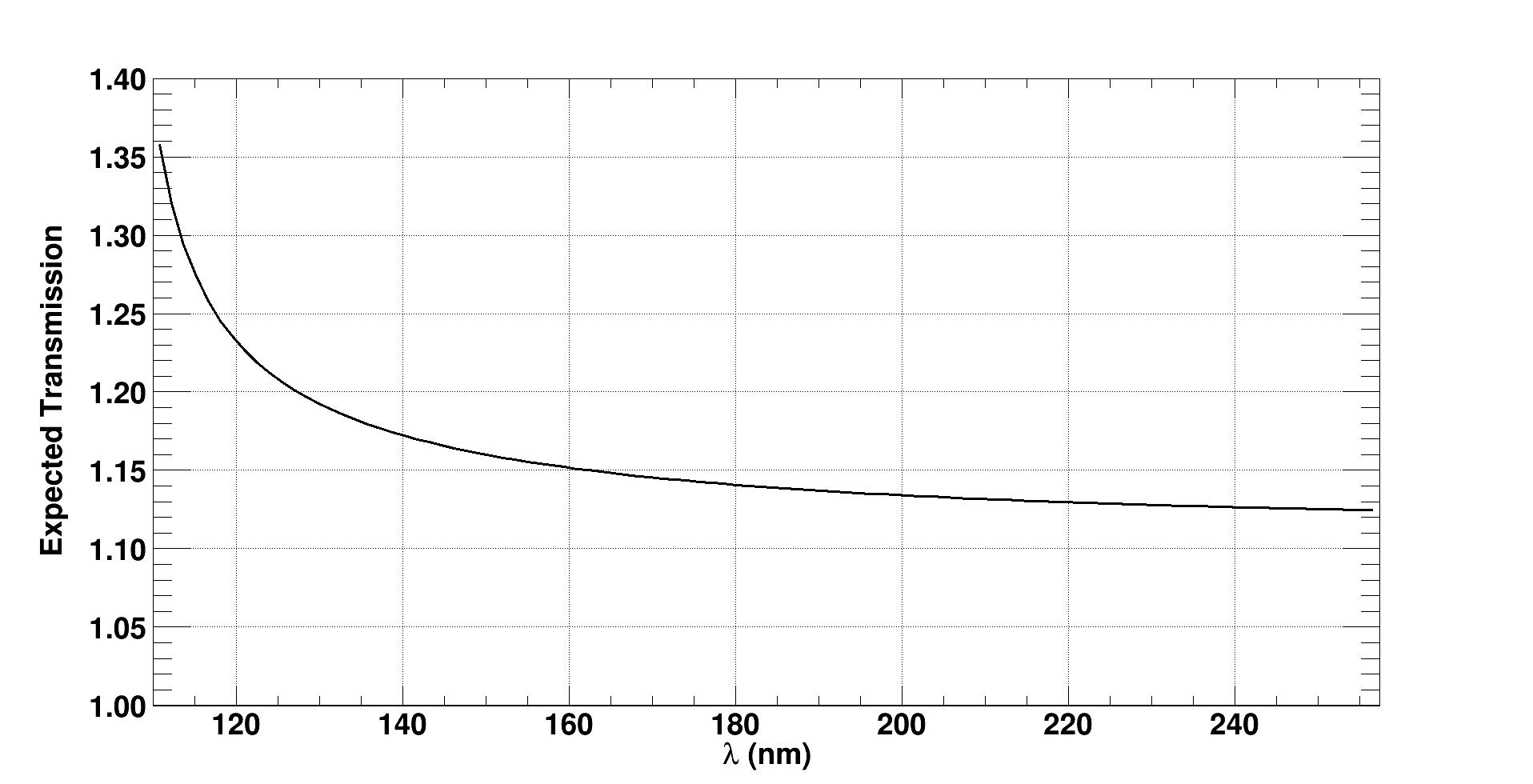}
 \caption{\textit{The expected transmission of liquid argon without any absorption or attenuation features is presented. The excessive transmission values are calculated using the mathematical product of equations 1 and 2 which describe the combination of the "Fresnel effect" (see figure\,7) at normal incidence of light and the enhanced transmission due to a light source with finite divergence. The geometry is selected according to the upper drawing in figure\,8.}}
 \label{fig:fresnel_fokussierung_kombiniert}
\end{figure}

The combination of both effects (Fresnel and divergence, Figs.\,\ref{fig:fresnel_senkrecht} and \ref{fig:fokussierungseffekt}, respectively) for our geometry with x=17.5\,cm, y=11.6\,cm, z=36.5\,cm is calculated in figure\,\ref{fig:fresnel_fokussierung_kombiniert} versus wavelength according to the mathematical product of equations \ref{eq:fresnel_senkrecht} and \ref{eq:strahlfokussierung}. 

\paragraph{3.3 Wavelength-resolved Transmission Measurements in the VIS and NIR Region}

\begin{figure}
 \centering
 \includegraphics[width=\columnwidth]{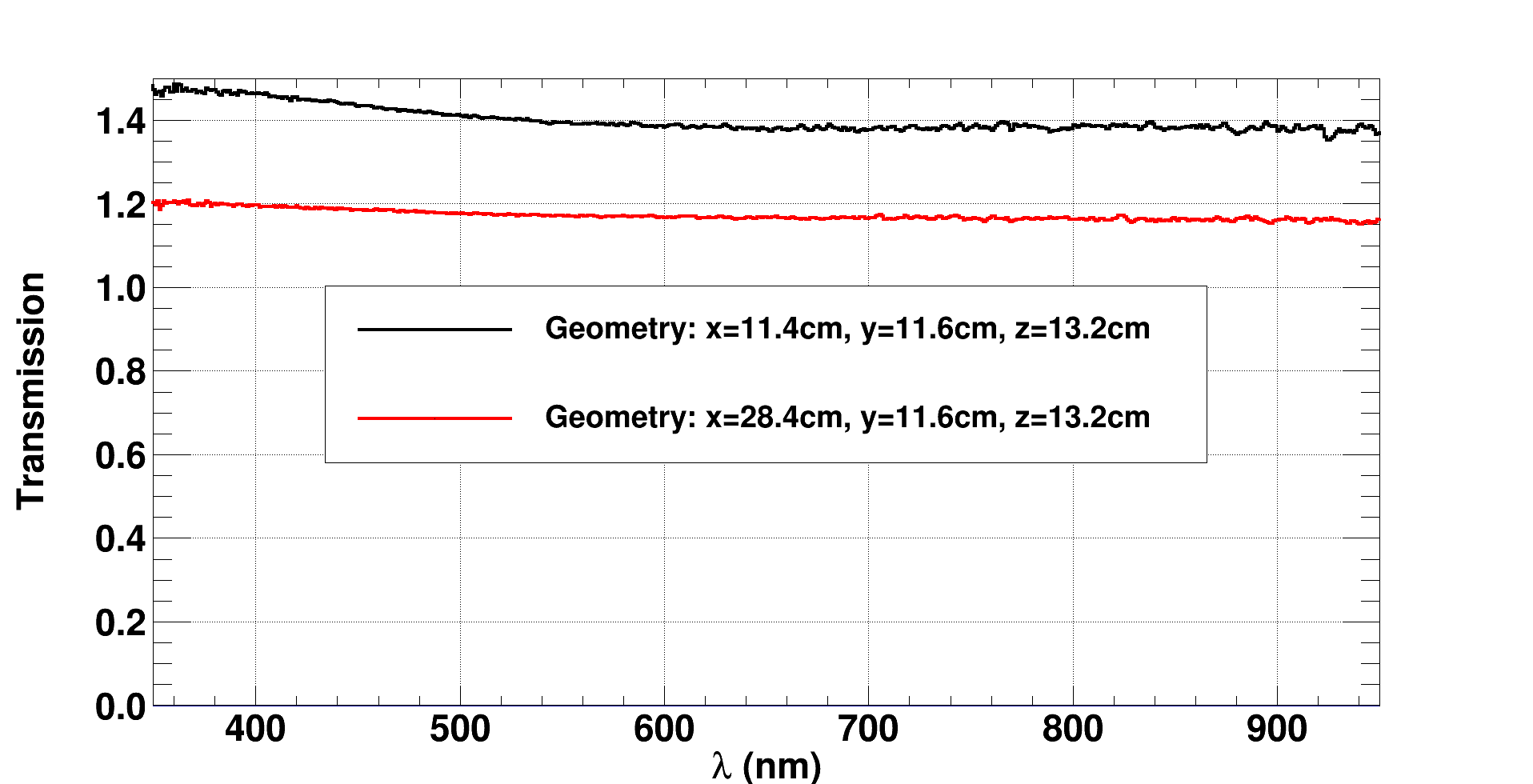}
 \caption{\textit{The measured transmission raw data of liquid argon in the visible and near-infrared region is presented for two different distances x. The light source was a halogen lamp, and the light was detected with a compact grating spectrometer with fiber optics coupling (Ocean Optics QE65000). The enhanced transmission values show that the effects described in subsections 3.1 and 3.2 do also appear in the visible region and are not due to some uncontrolled behaviour in the VUV.}}
 \label{fig:transmission_lar_oceanOptics}
\end{figure}

To verify that the geometric effect is not due to some uncontrolled behavior of the VUV setup we have extended the transmission measurements into the ultraviolet, visible and near-infrared wavelength regions using a halogen lamp and a small grating spectrometer with fiber optics input as detector. Figure\,\ref{fig:transmission_lar_oceanOptics} shows the transmission of pure liquid argon in the visible and near-infrared region. A decreasing source distance leads to increased transmission values. Thus, the measured transmission depends on the distance between light source and sample cell, which could be verified also in the ultraviolet, visible, and near-infrared wavelength ranges. 

\paragraph{3.4 Beam Profile Measurements with a He-Ne Laser}
\label{subsec:He_Ne_laser}

\begin{figure}
 \centering
 \includegraphics[width=\columnwidth]{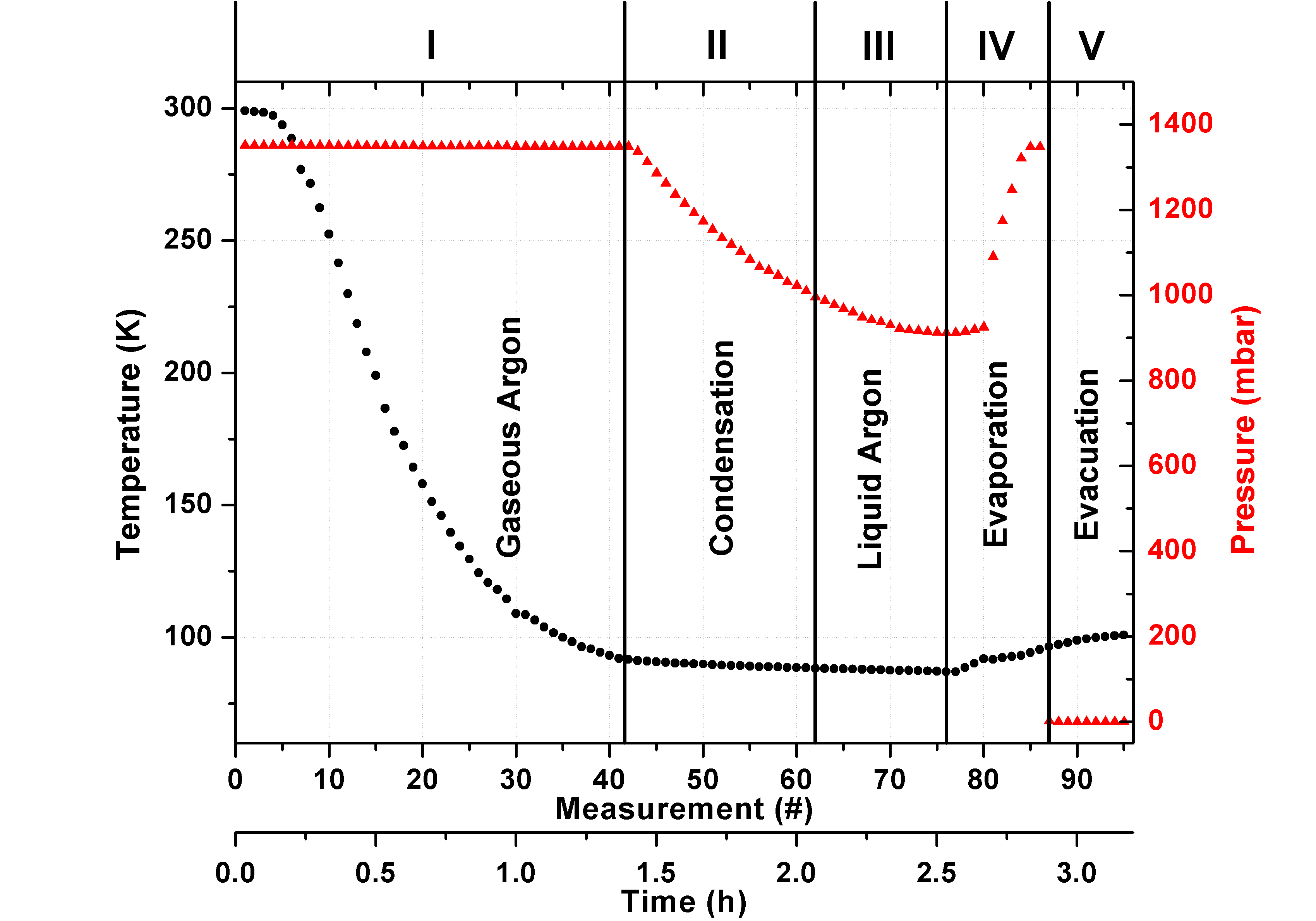}
 \caption{\textit{The temperature (left ordinate, black dots) and the pressure (right ordinate, red triangles, colour online) in the inner cell versus the measurement number (upper x axis) of the CCD camera (Atik 383Lc+) are shown. The CCD camera was programmed to take a measurement every two minutes with an exposure time of 200\,ms. The measurement campaign was started when the dewar was filled with liquid nitrogen. The lower x-axis shows the time in hours elapsed since filling the dewar with liquid nitrogen. The argon gas cooled down (phase I), condensed into the inner cell (phase II and III), was evaporated (phase IV) and evacuated (phase V) again. On the upper part of the figure the different phases of the campaign are indicated. The important phases are phase III when the laser beam went completely through liquid argon and phase V when the inner cell was evacuated.}}
 \label{fig:ATIK_Druck_Temp}
\end{figure}

In addition to the measurements in the UV, VIS and NIR range, it was also tested that even a 632.8\,nm He-Ne laser beam with a small divergence of $\sim0.5$\,mrad is measurably intensity enhanced by the Fresnel and geometry effects. A second aspect of this test was to study how (strongly) the beam is deflected during the condensation of liquid argon into the sample cell. The beam profile was measured using a CCD camera (Atik 383Lc+). The CCD chip (Kodak KAF-8300) has 3326 x 2504 active pixels on an area of 17.96 x 13.52\,mm which leads to a resolution of 5.4\,$\mu$m per pixel. 
The distances in the experimental setup in analogy to figure\,\ref{fig:fokussierung_schematisch} were: x=54\,cm, y=11.6\,cm, z=14.6\,cm. Note that x=54\,cm does not correspond to the distance between the laser output window and the liquid argon volume (27\,cm). A detailed investigation of the expansion of the laser beam profile with distance shows that the laser can be described (at large distances) as a point-shaped light source with a divergence of $\sim0.5$\,mrad with its origin $\sim27$\,cm within the laser resonator. Therefore, the relevant distance x in this experiment can be calculated as 54\,cm.  To adjust the intensity of the laser beam to the dynamic range of the camera an attenuator and two polarizing filters have been used. The measurement campaign was started when the dewar was filled with nitrogen and the camera was programmed to take a picture of the beam spot every two minutes with an exposure time of 200\,ms. Figure\,\ref{fig:ATIK_Druck_Temp} depicts the temperature (left y-axis) and the pressure (right y-axis) in the inner cell during the measurement campaign. The x-axis exhibits time and measurement number of the CCD camera, and the upper part shows the different phases of the experiment which will be explained in the following:

\begin{description}
 \item[Phase I:] The measurement campaign starts with filling the dewar with liquid nitrogen. The inner cell is at room temperature and filled with 1350\,mbar gaseous argon. The dewar filled with liquid nitrogen leads to a decreasing temperature of the inner cell whereas the pressure remains essentially constant.
 
 \item[Phase II:] At a temperature of $\sim 90$\,K the argon gas condenses into the cooled inner cell. Therefore, the temperature stays rather constant and the pressure begins to decrease.
 
 \item[Phase III:] At a pressure of $\sim 1000$\,mbar the laser beam goes completely through liquid argon, i.e., the inner cell is filled to $\sim 60$\,\% liquid argon. Therefore, in the following discussion, phase III is called the liquid phase although the condensation process is still in progress leading to further pressure reduction in the gas phase above the liquefied argon. At the end of phase III the sample cell is completely filled with liquid argon.
 
 \item[Phase IV:] At the beginning of phase IV the heating resistor is turned on and the temperature increases whereas the pressure increases only marginally. The liquid in the inner cell becomes an overheated fluid and at a certain temperature the pressure is drastically increased due to the liquid argon starting to boil.
 
 \item[Phase V:] When all of the liquid is evaporated the vacuum pump is turned on and the inner cell is evacuated. Therefore, the pressure drops to zero while the inner cell is still heated due to the heating resistor. In the following discussion phase V is called the vacuum phase. 
\end{description}

\begin{figure}
 \centering
 \includegraphics[width=\columnwidth]{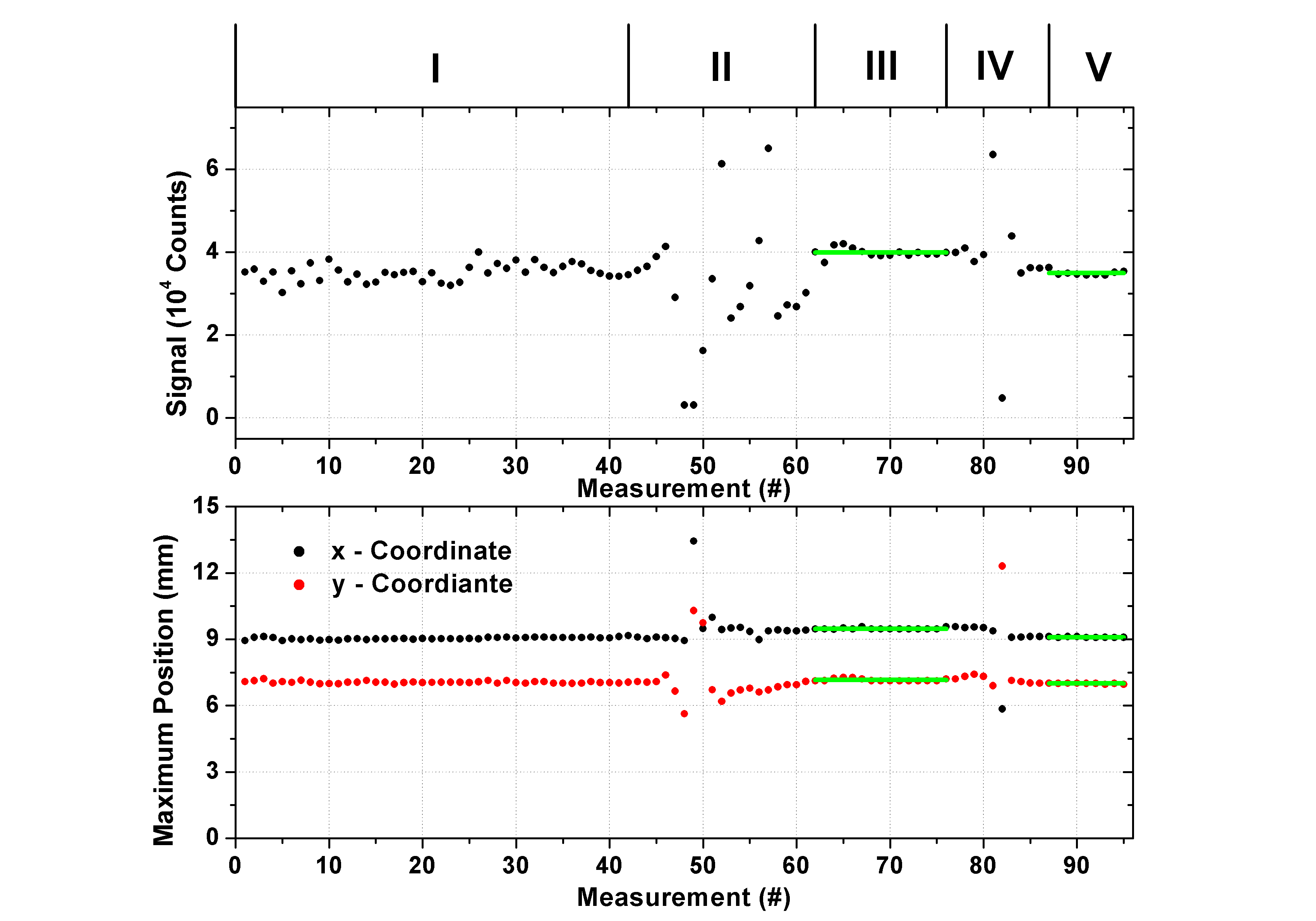}
 \caption{\textit{The signal in the maximum of the laser beam profile (upper panel) and the coordinates of the maximum position (lower panel) are shown. On top of the upper panel the different phases in the measurement campaign are indicated in analogy to figure\,12. When the inner cell is filled with liquid argon (phase III) the signal is $\sim14$\,\% enhanced compared to the evacuated inner cell (phase V). The horizontal green (colour online) lines denote the mean values in phases III and V. The positions of the maximum (lower panel) are measured relative to the lower left corner of the CCD chip. The mean values (green lines) of the x (black points) and y coordinates (red points) indicate that the signal is slightly shifted towards the upper right corner when the inner cell is filled with liquid argon (phase III) compared to the evacuated inner cell (phase V). See text for details. The mean values are summarized in table 1.}}
 \label{fig:ATIK_Maximum_Position}
\end{figure}

Phases III and V are the important ones which have to be compared, since these phases correspond to the cases when the sample cell is filled with liquid argon and evacuated, respectively. Under these conditions the wavelength-resolved measurement and reference spectra have been obtained which lead to the transmission spectra presented in Figs.\,\ref{fig:transmission_lar} and \ref{fig:transmission_lar_oceanOptics}. The beam profiles of the He-Ne laser in each picture were background corrected and fitted with a Gaussian profile in horizontal and vertical direction relative to the CCD chip orientation. The upper panel in figure\,\ref{fig:ATIK_Maximum_Position} shows the measured signal in the peak of the beam spot. The different phases in analogy to figure\,\ref{fig:ATIK_Druck_Temp} are indicated on the top of the figure. A comparison of the signal heights between phase III and phase V shows an increased signal in the liquid which can be explained by the "Fresnel effect" and the finite divergence of the laser beam in analogy to the transmission values above 1.0 in Figs.\,\ref{fig:transmission_lar} and \ref{fig:transmission_lar_oceanOptics}. It is also interesting to note that in phases II and IV there are huge signal variations due to instabilities of the gas-liquid boundary leading to chaotic light scattering. A further feature which can be observed is that during the cool-down in phase I the variation of the signal is much stronger than in phases III and V. The lower panel of figure\,\ref{fig:ATIK_Maximum_Position} shows the x- (black points) and y-coordinates (red points) of the position of the maximum of the beam profile. The origin of the coordinate system is in the lower left corner of the CCD chip looking in beam direction. A comparison between phases III and V shows a small deflection of the beam in the liquid to the upper right corner. This small deflection can be traced back to a not completely perpendicular alignment of the MgF$_{2}$ windows of the inner cell compared to the laser beam direction.

\begin{figure}
 \centering
 \includegraphics[width=\columnwidth]{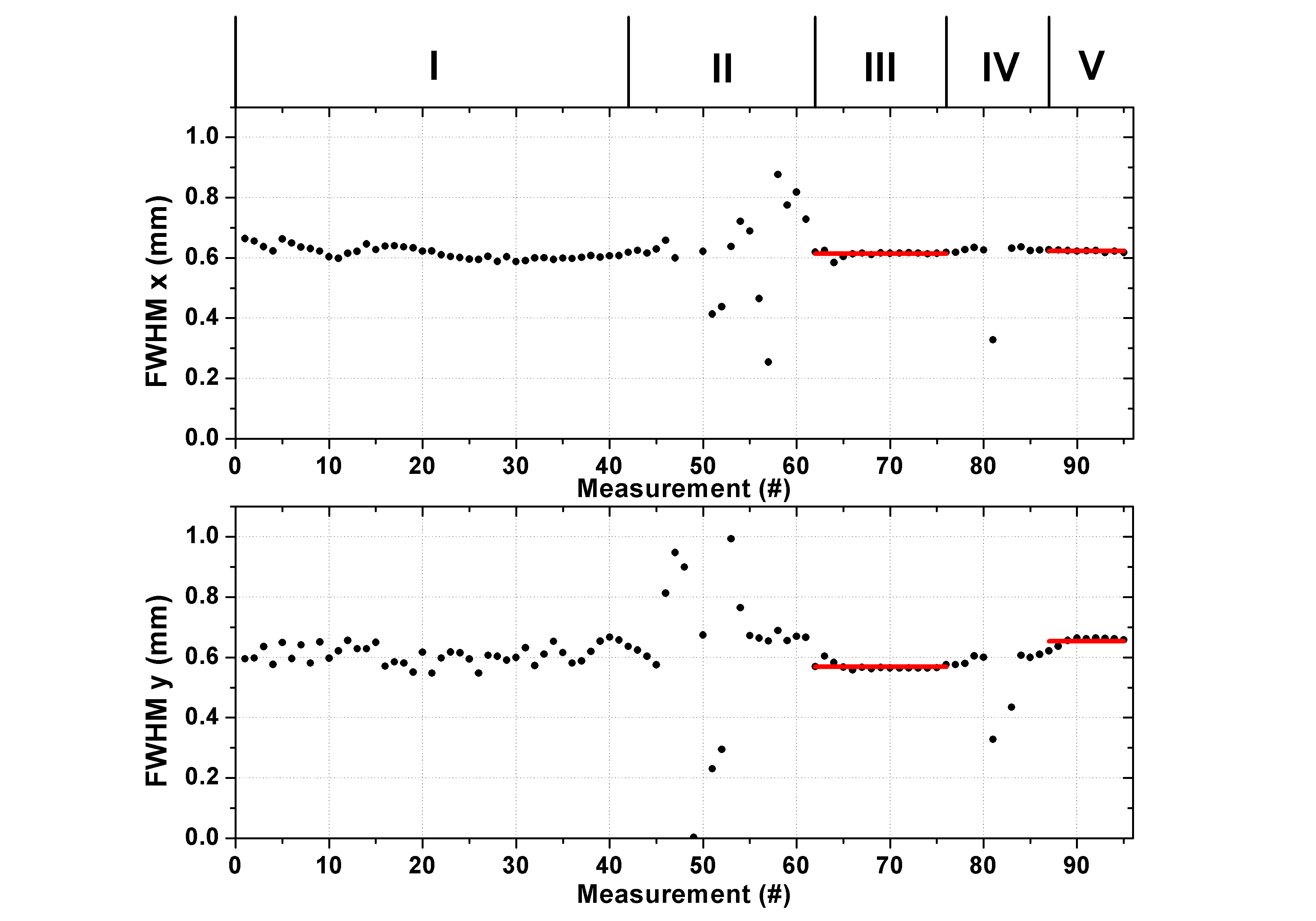}
 \caption{\textit{The full width at half maximum (FWHM) values of the laser beam in horizontal (upper panel) and vertical (lower panel) direction are shown versus measurement number. The laser beam was fitted in the peak with a gaussian profile in horizontal (x) and vertical (y) direction. The red (colour online) lines denote the mean values in phases III and V when the inner cell is filled with liquid argon and evacuated, respectively. Filling the inner cell with liquid argon leads to a reduction of the FWHM by $\sim 1.5$\,\% in horizontal and $\sim13$\,\% in vertical direction compared to the evacuated inner cell due to a finite divergence of the laser beam ($\sim0.5$\,mrad).}}
 \label{fig:ATIK_FWHM}
\end{figure}

Figure\,\ref{fig:ATIK_FWHM} shows the full width at half maximum (FWHM) values of the Gaussian fits to the beam profiles in x- (upper panel) and y- (lower panel) direction. Again, the different phases in analogy to figure\,\ref{fig:ATIK_Druck_Temp} are indicated on the top of the upper panel. A comparison between phases III and V shows that in both cases the beam profile is narrowed when the sample cell is filled with liquid argon. This demonstrates that the "finite divergence effect" can be measured even with a light source with a very narrow beam divergence ($ \sim0.5$\,mrad) like the He-Ne laser used in this experiment. So far we have no explanation why the narrowing of the beam profile is stronger in y-direction (vertical) compared to the x-direction (horizontal). It is also interesting to note that in analogy to phase I in figure\,\ref{fig:ATIK_Maximum_Position} also here in phase I during the cool-down the variation of the FWHM values is larger compared to phases III and V. A comparison of phase I between upper and lower panel shows that the variation is stronger in the vertical direction compared to the horizontal direction. Again, phases II and IV show the data points when the gas-liquid phase-transition goes through the laser beam which leads to strong deflections and uncontrolled behaviour of the beam profile which causes a large variation of the FWHM values. 

The horizontal lines in Figs.\,\ref{fig:ATIK_Maximum_Position} and \ref{fig:ATIK_FWHM} denote the mean values in phase III and phase V which correspond to the inner cell filled with liquid argon and the evacuated inner cell, respectively. In table\,1 these mean values are summarized and will be discussed in the following. 

\begin{table*}
\label{tab:ATIK_Fit_Values}
\smallskip
\centering
\begin{tabular}{||l||c|c|c|}
\hline
\hline
                        & Liquid Argon        & Vacuum            & Relative Change\\
\hline
\hline
Signal (Counts)         &$39886\pm286$        & $34963\pm192$     & $1.141\pm0.010$\\
\hline
Maximum Position x (mm) &$9.478\pm0.010$      & $9.093\pm0.007$   & $1.042\pm0.001$\\
\hline
Maximum Position y (mm) &$7.158\pm0.015$      & $6.998\pm0.007$   & $1.023\pm0.002$\\
\hline
FWHM x (mm)             &$0.613\pm0.002$      & $0.622\pm0.001$   & $0.985\pm0.004$\\
\hline
FWHM y (mm)             &$0.569\pm0.003$      & $0.654\pm0.005$   & $0.871\pm0.008$\\
\hline
\end{tabular}
\caption{\textit{Mean values for the peak signals, maximum positions and values for the distribution width (FWHM), extracted from figures 13 and 14. The columns correspond to the cell filled with pure liquid argon (phase III) and evacuated (phase V), respectively. The last column shows the relative changes of the parameters in respect to the empty cell (phase V). Errors of the mean are listed.}}
\end{table*}

The last column of table\,1 shows the relative change of all measured parameters from the inner cell filled with liquid argon compared to the evacuated inner cell. The signal in the peak of the laser beam spot is enhanced by $\sim14$\,\% (compared to the evacuated inner cell) which can be attributed to a combination of the "Fresnel effect" for perpendicular incidence of light and the "finite divergence effect" discussed in subsections 3.1 and 3.2. 

Here it is interesting to compare the results from the He-Ne laser with the results from the halogen lamp at the wavelength of the laser at 632.8\,nm. From figure\,\ref{fig:transmission_lar_oceanOptics} an enhanced transmission of 16.6\,\% for the long source distance (x=28.4\,cm, in figure\,\ref{fig:fokussierung_schematisch}) and 37.6\,\% for the short source distance (x=11.4\,cm, in figure\,\ref{fig:fokussierung_schematisch}) is found at 632.8\,nm (the wavelength of the laser). In table\,2 the enhanced transmission values above 1.0 for a laser wavelength of 632.8\,nm are summarized and compared with predictions according to equations \ref{eq:fresnel_senkrecht} and \ref{eq:strahlfokussierung}. A comparison of the transmission values of the halogen lamp (assuming the impact of the "Fresnel effect" is equal in both cases) shows that the "finite divergence effect" leads to increasing transmission values with decreasing source distances x. The He-Ne laser which has the most narrow beam profile ($\sim 0.5$\,mrad) and the largest distance (x=54.0\,cm) shows the smallest transmission above a value of 1.0. A comparison of measured transmissions with expected transmissions (compare lines three and six in table\,2) in the case of the He-Ne laser as well as the halogen lamp shows that eqs. \ref{eq:fresnel_senkrecht} and \ref{eq:strahlfokussierung} describe the trend qualitatively correct. However, a reliable quantitative description can only be given if measurements of the refractive index of liquid argon as well as MgF$_{2}$ are available\footnote{The calculations of the expected transmissions at the laser wavelength rely on extrapolated refractive indices of liquid argon (n=1.22) as well as MgF$_{2}$ (n=1.37). The model for the wavelength-dependent refractive index of liquid argon is adapted from ref.\,\cite{Bideau-Mehu} and the model for the wavelength-dependent refractive index of MgF$_{2}$ is adapted from ref.\,\cite{laporte}.}. The product of equations \ref{eq:fresnel_senkrecht} and \ref{eq:strahlfokussierung} describe the expected transmission above a value of 1.0 quite well for the He-Ne laser (40\,\% disagreement\footnote{A refractive index of liquid argon of n=1.38 would lead to a perfect agreement of measurement and prediction.} compared to the measurement) and the halogen lamp positioned at the long source distance (x=28.4\,cm, 27\,\% disagreement\footnote{A refractive index of liquid argon of n=1.30 would lead to a perfect agreement of measurement and prediction.} compared to the measurement). The measurement with the halogen lamp positioned at the short source distance (x=11.4\,cm), however, exhibits a significantly higher transmission than predicted by the product of equations \ref{eq:fresnel_senkrecht} and \ref{eq:strahlfokussierung} (factor of $\sim$2 disagreement\footnote{A refractive index of liquid argon of n=1.77 would lead to a perfect agreement of measurement and prediction.}). So far we have no reliable explanation for this discrepancy. During the measurements with the halogen lamp it turned out that the orientation of the fiber optics relative to the halogen lamp had a strong influence on the measured signal. A slight deflection (see below) of the laser beam when the cell is filled with liquid argon has been measured. This deflection could lead to a better coupling of the transmitted light into the fiber optics when the cell is filled with liquid argon and, consequently, to an artificially increased transmission. However, so far, a further unresolved systematic effect cannot be excluded.

\begin{table*}
\label{tab:Signal_Comparison_Laser_Halogen}
\smallskip
\centering
\begin{tabular}{||m{0.29\textwidth}||c|c|c|}
\hline
\hline
Light Source                                           & He-Ne Laser         & Halogen lamp          & Halogen lamp\\
\hline
\hline
Distances x - y - z (cm)                               &$54 - 11.6 - 14.6$   & $28.4 - 11.6 - 13.2$  & $11.4 - 11.6 - 13.2$\\
\hline
Measured Transmission                                  &$1.141$              & $1.166$               & $1.376$\\
\hline
"Fresnel effect"                                       &$1.044$              & $1.044$               & $1.044$\\
\hline
"Finite divergence effect"                             &$1.054$              & $1.083$               & $1.126$ \\
\hline
"Fresnel" and "finite divergence effect" combined      &$1.100$              & $1.131$               & $1.176$\\
\hline
\end{tabular}
\caption{\textit{Results on the increased transmission ($>1.0$) at a wavelength of 632.8\,nm (laser wavelength) obtained with the He-Ne laser (see figure 13 upper panel and table 1, respectively) and the halogen lamp (see figure\,11 at 632.8\,nm). Expected values according to equations 1 and 2 at the laser wavelength are also presented. The second line shows the distances x,y and z according to the geometry in figure 8. The third line presents the measured transmissions above a value of 1.0 for the He-Ne laser and for the halogen lamp. The fourth line shows the expected transmission above a value of 1.0 due to the "Fresnel effect" for normal incidence of light according to equation 1 at the laser wavelength. The fifth line shows the expected transmission values above 1.0 due to the "finite divergence effect" according to equation 2 at the laser wavelength. The sixth line shows the expected transmissions above 1.0 from the combination of the "Fresnel" and the "finite divergence effect", i.e., the mathematical product of equations 1 and 2 at the laser wavelength. Note that the values for the refractive indices of liquid argon (n=1.22) and MgF$_{2}$ (n=1.37) at a wavelength of 632.8\,nm are extrapolated using the wavelength-dependent models in refs.\,\cite{Bideau-Mehu} and \cite{laporte} since no measurements are available in that wavelength region.}}
\end{table*}

The changes of the coordinates of the maximum of the beam profile are a shift of $\sim0.4$\,mm in positive x- direction and a shift of $\sim0.2$\,mm in positive y-direction (compare lines two and three in table 1). That means the maximum position of the beam profile is shifted a little bit towards the upper right corner compared to the position when the sample cell is evacuated. This shift can be explained by a not perfectly perpendicularly aligned geometry of the laser beam and the inner cell, i.e., the angle between the MgF$_{2}$ windows and the laser beam is not exactly 90\textdegree. A simple estimation based on the upper panel of figure\,\ref{fig:fokussierung_schematisch} leads to a misalignment of the optical axis of the inner cell relative to the laser beam by $\sim4$\,mrad in x and $\sim1.7$\,mrad in y-direction.

The widths (FWHM) of the beam profiles are decreased in liquid argon compared to the evacuated inner cell to $\sim98$\,\% in horizontal and $\sim87$\,\% in vertical direction (compare lines five and six in table 1). Due to the high spatial resolution of the CCD camera the "finite divergence effect" could directly be measured even with a light source with a very narrow beam profile like a He-Ne laser. So far we have no explanation for the $\sim8$ times stronger beam narrowing in vertical direction compared to the horizontal direction. However, a density related effect due to a temperature gradient could lead to a variation of the refractive index in vertical direction.

\section{Quantitative Test of the "Fogging Correction" and Estimate of the Systematic Error}
\label{sec:fogging_correction}

\begin{figure}
 \centering
 \includegraphics[width=\columnwidth]{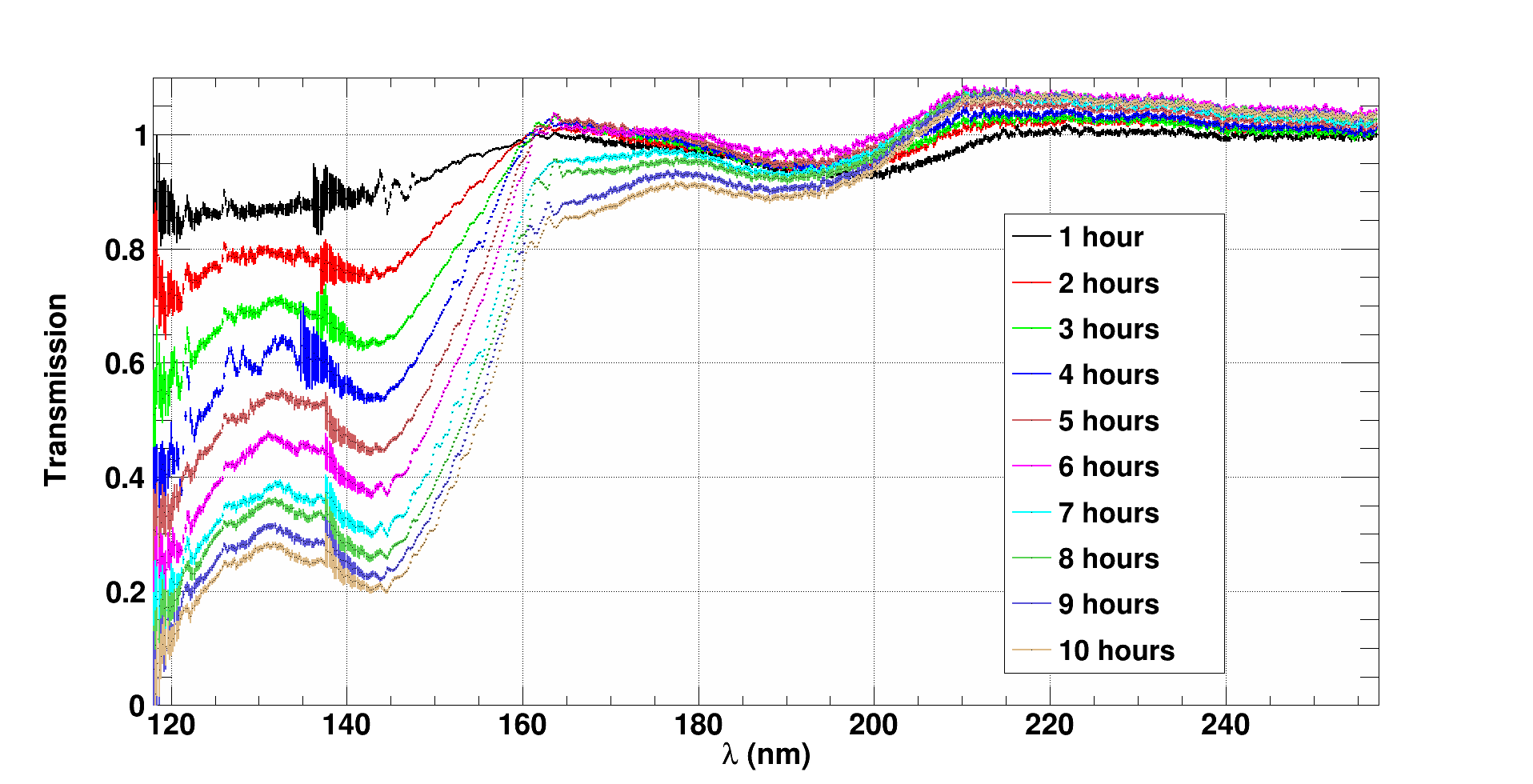}
  \caption{\textit{The transmission of the cooled and empty (evacuated) inner cell for different cooling times is presented. A reference spectrum has been measured when the inner cell was at room temperature. Every hour after filling the dewar with liquid nitrogen a spectrum has been taken and the transmission has been calculated. In the measurements the temperature of liquid argon is reached after about 2 hours. Below 160\,nm the alteration of the transmission is severe, down to a transmission of approximately 20\% at 142\,nm after 10 hours cooling time. This can be attributed to the formation of ice on the cold  MgF$_{2}$ windows from water vapor in the insulation vacuum of the outer cell. Water frozen to ice can have (dependent on the substrate temperature) strong absorption bands at 142\,nm \cite{onaka_uv_ice}. In the outer cell a pressure in the range of $10^{-7}$mbar has been reached during the measurements. This shows that the inner cell is basically acting as a cold trap. Note the enhanced transmission above 210\,nm. It could be due to two effects which lead to this enhanced transmission values. Firstly, the coating of the cryogenic surfaces of the MgF$_{2}$ windows could lead to a deposit which has a refractive index between 1.0 and that of MgF$_{2}$ which leads to an enhanced transmission (optical coating). Secondly, this observation could be attributed to resonance fluorescence of the components frozen to the cold MgF$_{2}$ windows.}}
 \label{fig:beschlag_wellenlaenge}
\end{figure}

\begin{figure}
 \centering
 \includegraphics[width=\columnwidth]{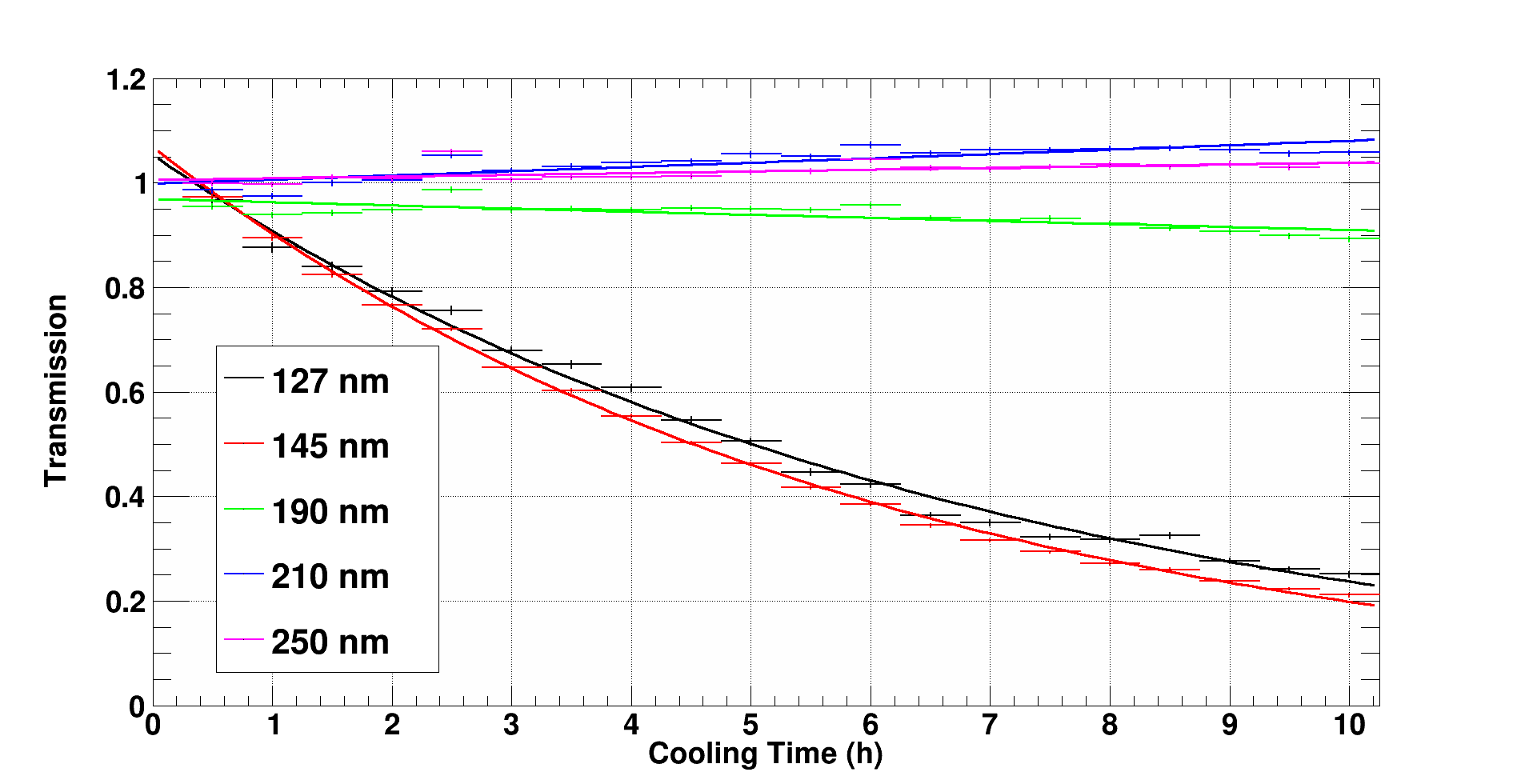}
 \caption{\textit{The time evolution of the transmission of the cold evacuated inner cell is presented for a set of selected wavelengths from figure\,15. To correct the measured transmission spectra of liquid argon the "fogging" of the cold MgF$_2$ windows has to be determined time-resolved. The solid lines show exponential fits to the time evolution of the transmission of the empty (evacuated) inner cell. The fit results obtained at each wavelength were used to correct the measured transmission spectra of liquid argon for this effect.}}
 \label{fig:beschlag_zeit}
\end{figure}

\begin{figure}
 \centering
 \includegraphics[width=\columnwidth]{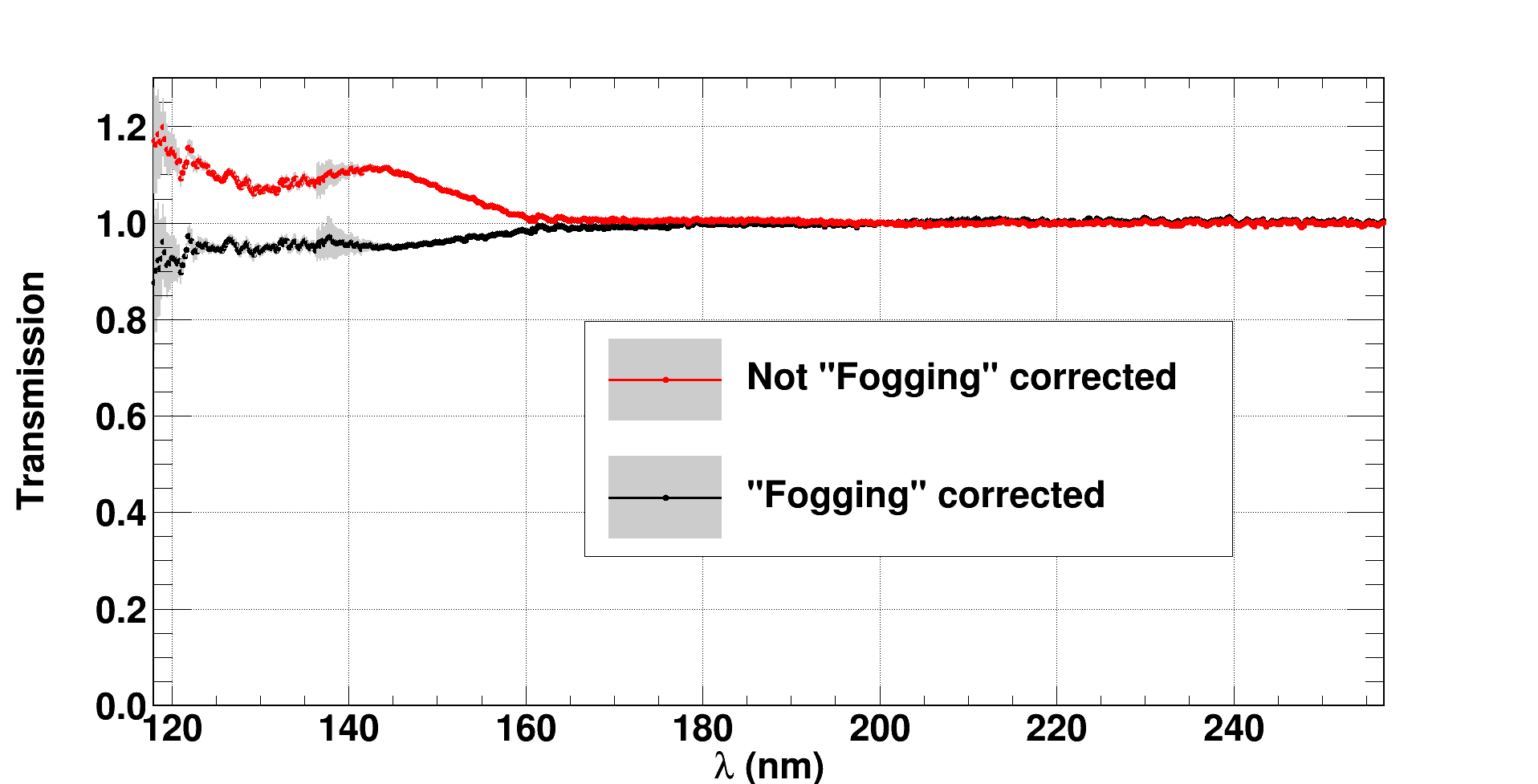}
 \caption{\textit{The measured transmission of the empty (evacuated) inner cell is presented (red (colour online) line with (gray) statistical error bars). The spectrum has been measured approximately 2 hours, the corresponding reference spectrum approximately 3 hours after filling the dewar with liquid nitrogen. These times roughly correspond to the measurement times when the transmission of liquid argon was measured. Taking the reference spectrum after the measurement of the inner cell filled with the liquid leads to transmission values above 1.0 due to the "fogging effect". The black line (with (gray) statistical error bars) shows the transmission spectrum after correcting the measured transmission at each wavelength using the fit values derived from figure\,16.}}
 \label{fig:beschlag_benchmark}
\end{figure}

All quantitative measurements of the attenuation of light in our setup have to rely on a good quantitative correction of an aspect which we have termed "fogging" in ref. \cite{Neumeier_epjc_2012} and described based on data shown in figure 4 in that reference: The sample cell is cooled down to approximately liquid nitrogen temperature during the experiments which takes about two hours for cooling, condensation of the gas, recording the spectra with liquid argon, evaporating the system and then taking the reference spectra with a cool, evacuated sample cell. During that time the windows of the sample cell are exposed to the vacuum of about 10$^{-7}$\,mbar in the CF100 cross piece which is evacuated by a turbo molecular pump. Cooling the inner cell leads to condensation of rest-gas components on the optical windows, presumably water vapor which is condensed to ice. Water, frozen to ice, is known to have a strong absorption band in the vacuum ultraviolet depending on the temperature of the ice \cite{onaka_uv_ice,blackman_ice_structures}. Figure\,\ref{fig:beschlag_wellenlaenge} shows the wavelength-dependent transmission of the empty inner cell for different times after filling the dewar with liquid nitrogen. Recording a spectrum typically took 6\,minutes. Figure\,\ref{fig:beschlag_zeit} displays the time dependence of the transmission for several wavelengths chosen from figure\,\ref{fig:beschlag_wellenlaenge}. To be able to correct for this effect the exact times of all the measurements were carefully recorded with respect to the starting time (dewar filling) of the cooling-down.
A recent cycle of measurements has now allowed us to test our "fogging correction". In the transmission measurements of liquid argon the cooling-down and the liquefication took approximately two hours and the following evaporation, evacuation and measurement of the reference spectrum has been performed typically three hours after filling the dewar with liquid nitrogen. Figure\,\ref{fig:beschlag_benchmark} shows the transmission of the evacuated and empty inner cell obtained from spectra taken at the typical measurement times relative to filling the dewar with liquid nitrogen. If no "fogging effect" would deteriorate the transmission of the MgF$_2$ windows the transmission should be 1.0 for all measured wavelengths. Due to the fact that the reference spectrum has been measured always after the spectrum with the inner cell filled with liquid argon (liquid argon could be evaporated faster than condensed into the inner cell) the "fogging" of the MgF$_2$ windows leads to an artificially decreased reference spectrum. Therefore, the calculated transmission spectra always show a wavelength-dependent transmission above a value of 1.0. This has a strong effect for wavelengths shorter than 160\,nm. In figure\,\ref{fig:beschlag_benchmark}, the black curve shows the transmission of the evacuated inner cell without a "fogging correction". The red curve shows the same data after applying the "fogging correction". The transmission values are corrected to $\sim95$\,\% for wavelengths shorter than 160\,nm and to 1.0 for wavelengths longer than 160\,nm. To account for this systematic uncertainity we have added the deviation from 1.0 as a systematic error to the data presented in section \ref{sec:Transmission_LAr}, since for a perfect "fogging correction" one would expect that the transmission of the empty sample cell is corrected to 1.0 independently of wavelength.

\section{Improvement of the Data Analysis Procedure}
  \label{sec:dataAnalysisMinorImprovement}
\begin{figure}
 \centering
 \includegraphics[width=\columnwidth]{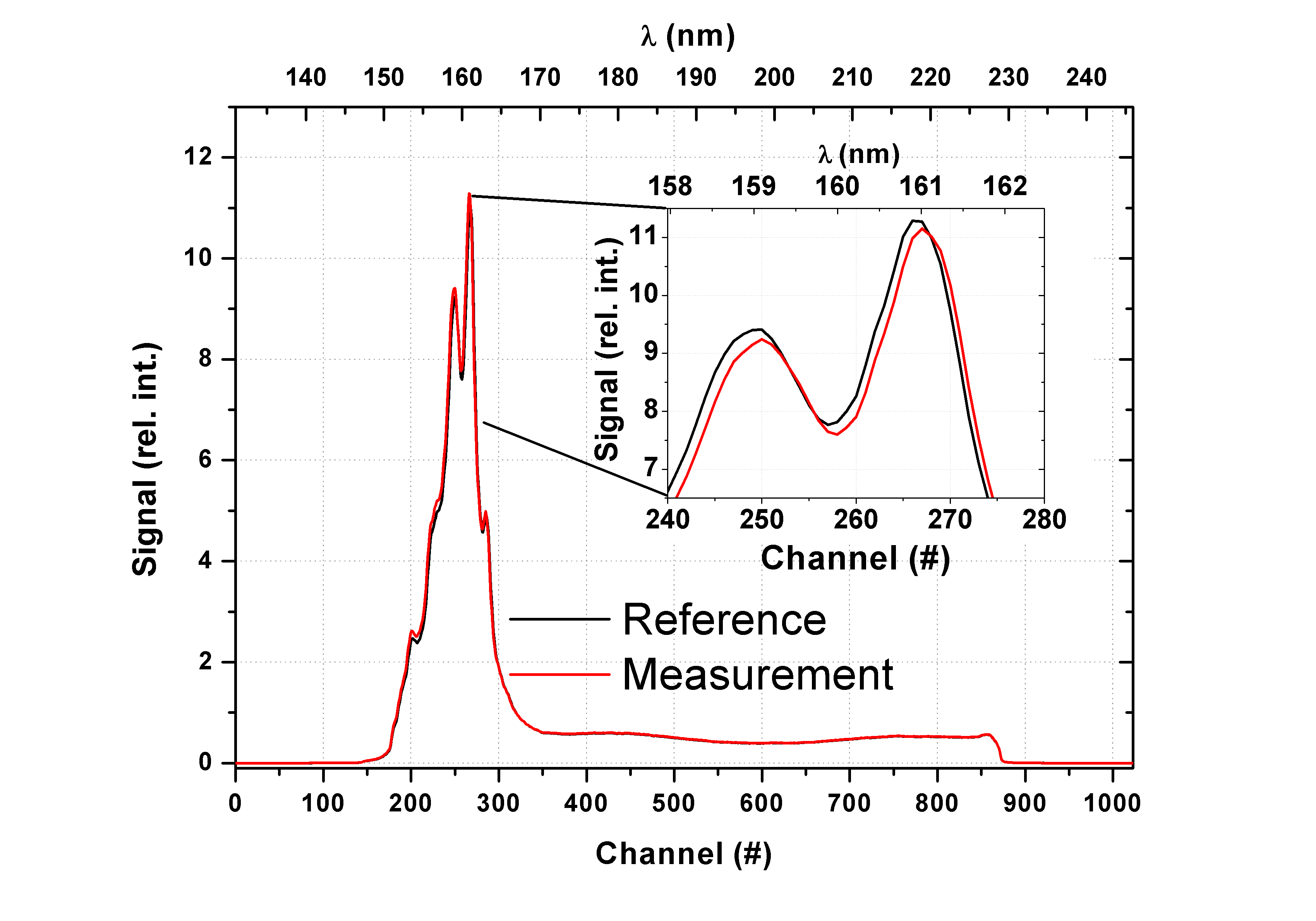}
 \caption{\textit{The emission spectra of the deuterium lamp through the evacuated and cooled inner cell are shown. The spectra were recorded with a central wavelength position of the monochromator at 180\,nm and cover the range $\sim(180\pm35)$\,nm. Below channel 180 and above channel 860 the detector is not sensitive to light since the diode array is blocked by the housing of the image intensifier. The red (colour online) curve shows a measurement spectrum and the black (colour online) curve the corresponding reference spectrum recorded approximately 1\,h after the measurement spectrum. The lower x-axis shows the number of channels and the upper x-axis the corresponding wavelength in nm. A detailed view of the most intense emission region is shown in the inset. A slight shift between measurement and reference spectra is visible. This shift is due to a not exactly equally adjusted central wavelength position of the monochromator between the measurement and the corresponding reference spectrum.}}
 \label{fig:D2_Emission_Messung_Referenz}
\end{figure}

\begin{figure}
 \centering
 \includegraphics[width=\columnwidth]{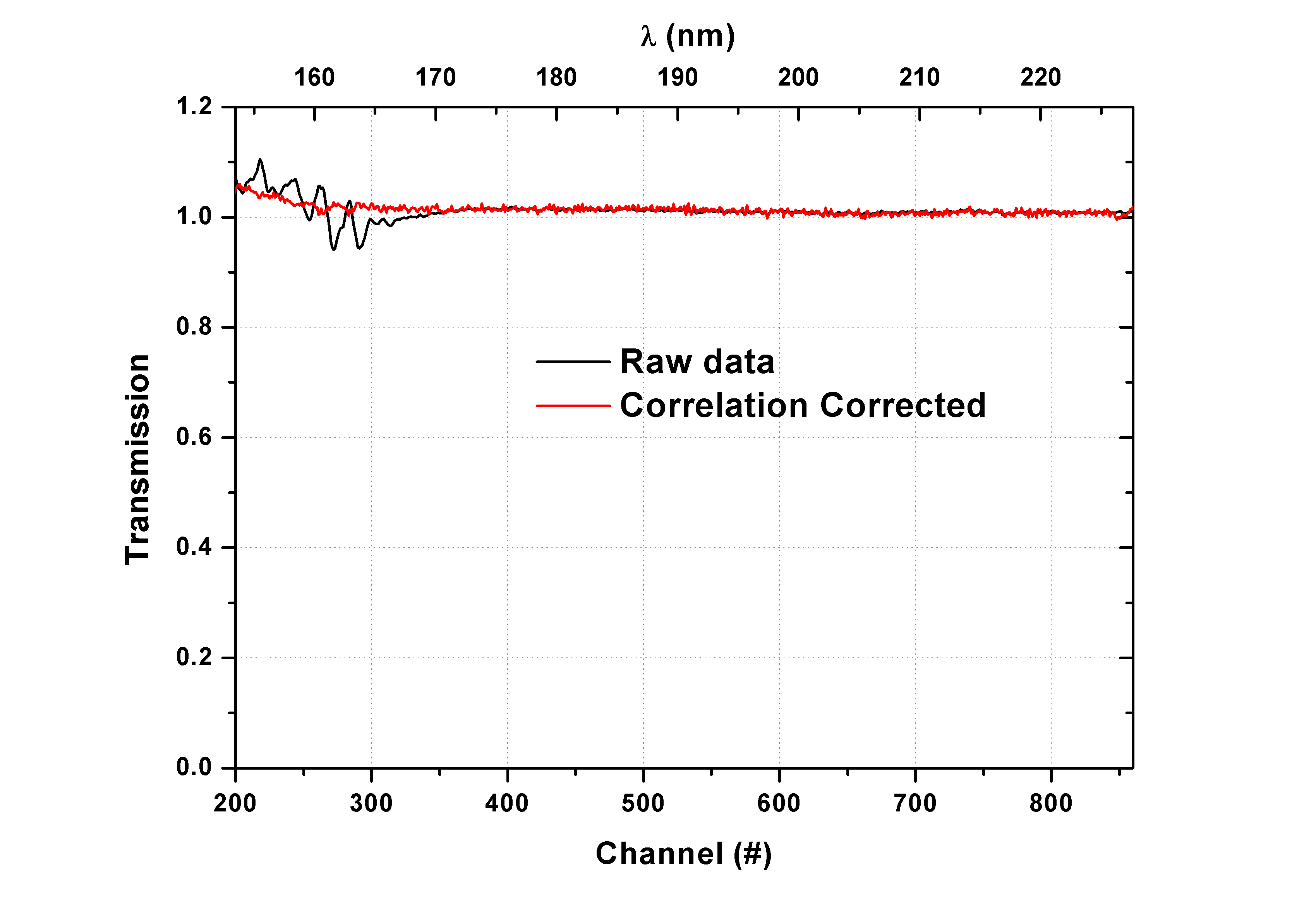}
  \caption{\textit{The transmission of the cooled and evacuated inner cell is shown. The black curve is obtained by a division of the data shown in figure\,18. A not exactly equal central wavelength position of the monochromator between measurement and reference spectrum leads to the oscillations of the transmission for wavelengths below $\sim165$\,nm. The red curve shows the transmission after the measurement spectrum has been shifted by 0.77 channels to the left compared to the reference spectrum. This shows that these small oscillations can be traced back to a systematic effect of the experimental setup (see text for details).}}
 \label{fig:Vergleich_Transmission_Rohdaten_Korrelationskorrektur}
\end{figure}

\begin{figure}
 \centering
 \includegraphics[width=\columnwidth]{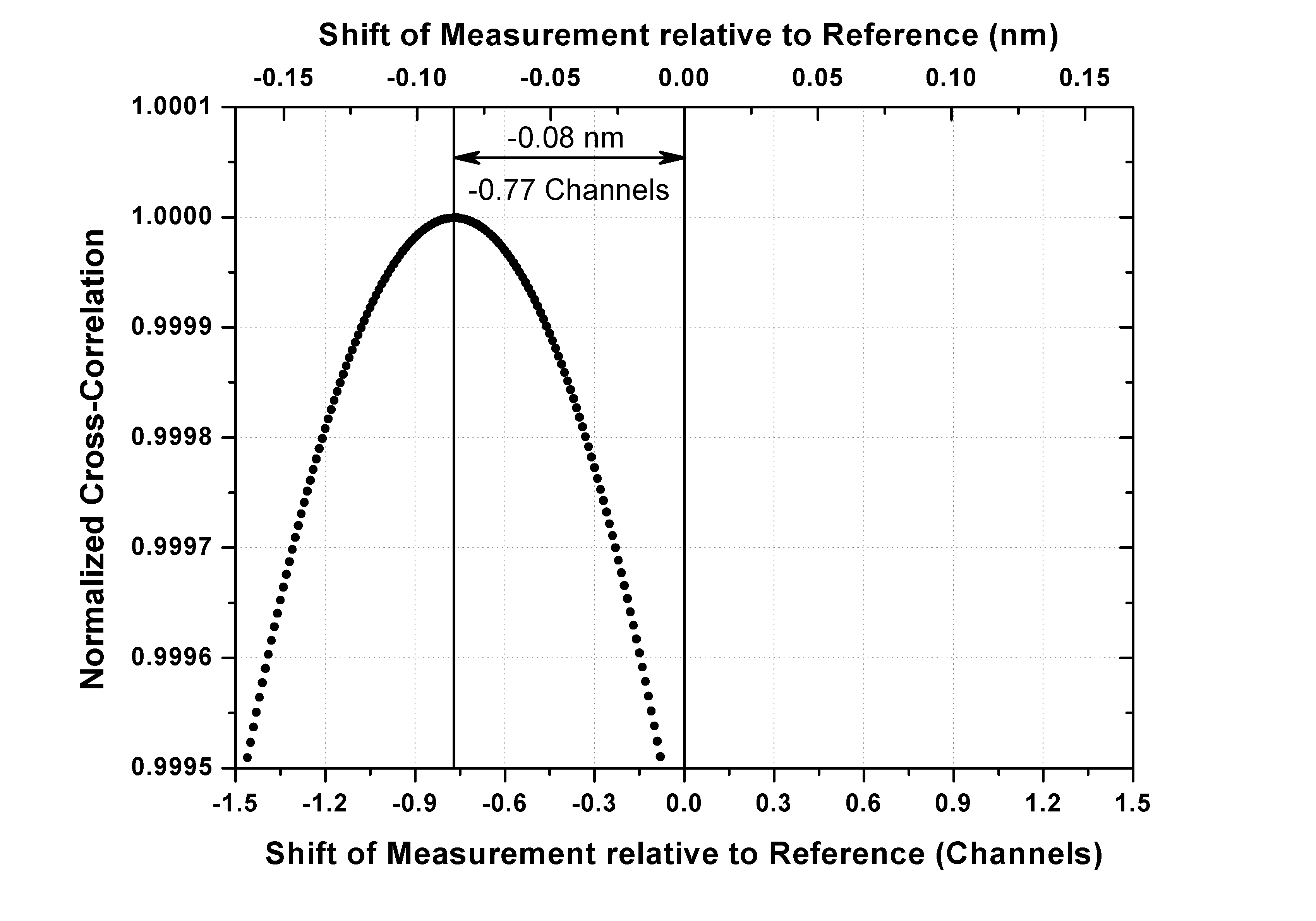}
 \caption{\textit{The cross-correlation between the measurement and the reference spectra from figure\,18 for different shifts in channels (lower x-axis) and nm (upper x-axis) of the measurement relative to the reference spectrum is shown. The maximum of the cross-correlation denotes the shift of the measurement spectrum necessary to be in maximal agreement with the reference spectrum.}}
 \label{fig:Kreuzkorrelation_Messung_Referenz}
\end{figure}

All the transmission measurements presented here are based on recording emission spectra of the deuterium lamp through the cell filled with a sample (measurement spectrum) and through the evacuated cell (reference spectrum). The measurement spectrum was divided by the reference spectrum to determine the transmission. The spectrometer covered a wavelength range of $\sim(\pm35)$\,nm around each setting of the central wavelength (see figure\,\ref{fig:D2_Emission_Messung_Referenz}). The central wavelength position of the spectrometer could be set manually by turning a fine pitch thread which adjusted the angle of the reflection grating in the monochromator to illuminate the image intensifier and, consequently, the diode detector array with different parts of the spectrum. Therefore, it was very important to adjust the spectrometer at exactly the same central wavelength positions for both the measurement spectrum as well as the reference spectrum. In each measurement campaign the central wavelength positions at 90, 120, 150, 180 and 210\,nm were investigated and after evaporating and evacuating the inner cell, reference spectra at exactly the same central wavelength positions had been measured. A small maladjustment of the central wavelength position of the monochromator between measurement and reference spectrum leads to a wavelength shift of $\lesssim 0.1$\,nm between them.  As an example, figure\,\ref{fig:D2_Emission_Messung_Referenz} displays a spectrum of the deuterium lamp through the evacuated (and cold) inner cell at 180\,nm central wavelength position of the monochromator. The red curve shows the measurement spectrum and the black curve the reference spectrum recorded $\sim1$\,h thereafter. The inset exhibits a detailed view of the spectrum at the most intense region (Lyman Band) of the deuterium lamp. A shift of $\lesssim 0.1$\,nm between measurement and reference spectra is visible. A simple division of the two spectra shown by the black curve in figure\,\ref{fig:Vergleich_Transmission_Rohdaten_Korrelationskorrektur} leads to an oscillating transmission below $\sim165$\,nm. The slight increase ($\sim5$\,\%) of the transmission is due to the above-explained "fogging effect" beginning from $\sim160$\,nm towards shorter wavelengths. From a spectroscopic point of view it was interesting to investigate whether these oscillations originate from liquid argon or if they come from a systematic effect. A detailed investigation (see figure\,\ref{fig:D2_Emission_Messung_Referenz} as an example) showed that in all the cases where these oscillations were observed in the transmission spectra a shift of $\lesssim 0.1$\,nm between measurement and reference spectrum was observed. For this reason an analysis procedure has been developed to shift the raw spectra in terms of channels to match each other. This analysis step has been introduced as first step before all other procedures. The analysis which follows this step has not been changed and is described in detail in ref.\,\cite{Neumeier_epjc_2012}. The shift which is necessary to bring measurement and reference spectra to an exact overlap is described as follows: Both spectra were interpolated using a cubic spline. Between each data point 100 new data points were created by interpolation. The next step was the calculation of the cross-correlation between measurement and reference spectra for different shifts. Figure\,\ref{fig:Kreuzkorrelation_Messung_Referenz} shows the normalized cross-correlation between measurement and reference spectra for different shifts of the measurement spectrum relative to the reference spectrum. The maximum of the cross-correlation is an indicator when the spectra are in maximal agreement. From this example one can deduce that the measurement spectrum has to be shifted by approximately 0.77 channels ($\sim0.08$\,nm) to the left compared to the reference spectrum to obtain a maximum in agreement. The red line in figure\,\ref{fig:Vergleich_Transmission_Rohdaten_Korrelationskorrektur} shows the 
transmission of the data set after a shift of the measurement spectrum by 0.77 channels ($\sim0.08$\,nm) to the left. The oscillations are strongly reduced and the "fogging effect" which manifests itself in an increased transmission towards shorter wavelengths is not affected by this procedure.

This improvement due to a fine adjustment of the measurement and reference spectrum is interesting concerning transmission measurements in a more general sense. This procedure is of particular importance due to the strong intensity variations in the emission spectrum of the deuterium lamp in this wavelength region (here the rotational structure in the Lyman-band). A light source with a smoother emission spectrum would strongly reduce this effect. Therefore, in general, light sources with a smooth emission spectrum or monochromatic light should be preferred for transmission measurements. In this case a broadband VUV source \cite{Dandl} based on electron-beam excited gaseous argon could be a good alternative.

\section{Optical Transmission of 11.6\,cm Pure Liquid Argon including all Corrections}
    \label{sec:Transmission_LAr}
\begin{figure}
 \centering
 \includegraphics[width=\columnwidth]{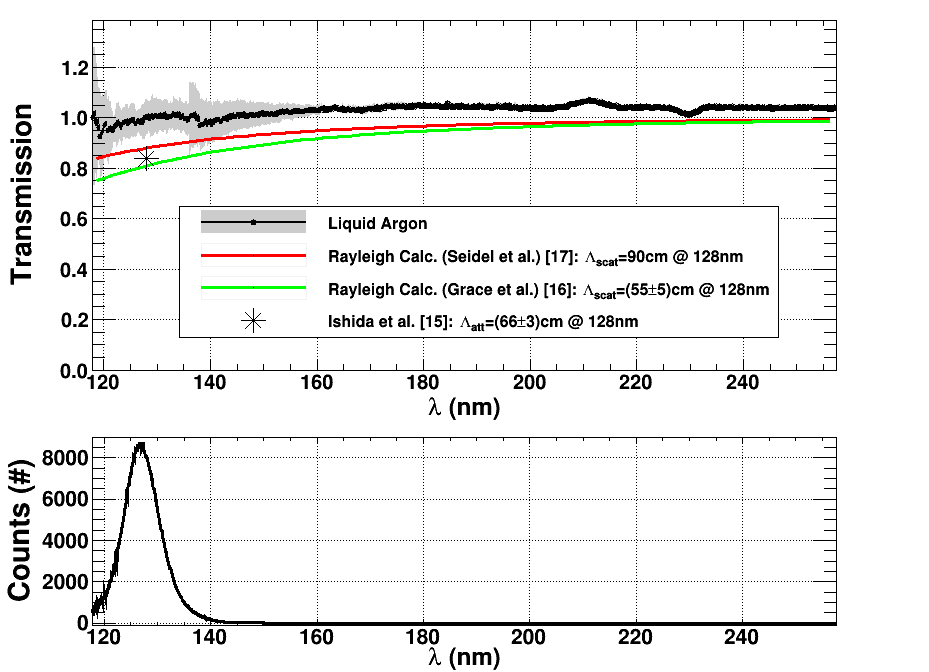}
 \caption{\textit{Upper panel: The corrected (Fresnel, finite divergence, "fogging" and cross-correlation) transmission of 11.6\,cm pure liquid argon is shown (upper panel, black solid curve). The gray error bars represent a linear sum of the statistical errors due to the different exposure times, the intensity of the deuterium lamp used and the systematical errors from the "fogging correction". Note that above 160\,nm the transmission is still $\sim4$\,\% too high (explanations see text). The red/green (colour online) solid curves show wavelength-dependent calculations of the expected transmission of 11.6\,cm pure liquid argon according to different values of a Rayleigh scattering length ($\Lambda_{scat}=90/(55\pm5)\,cm$ at 128\,nm, see refs.\,\cite{Rayleigh_Seidel,Grace_Nikkel}). The black asterisk shows, for comparison, the calculated value of the expected transmission of 11.6\,cm liquid argon for the attenuation length adapted from ref.\,\cite{Ishida} ($\Lambda_{att}=(66\pm3)\,cm$ at 128\,nm). Lower panel: For comparison the electron-beam induced emission of liquid argon is presented (adapted from ref.\,\cite{Heindl_epl}.)}}
 \label{fig:Transmission_Emission_LAr_Fresnel_Divergence_corrected}
\end{figure}

The main goal of this work is to measure the attenuation length of pure liquid argon for its own scintillation light. This is interesting in terms of scintillation-detector development since the detector volumes tend to increase and, therefore, also the optical path lengths for the scintillation light until it is detected. The upper panel in figure\,\ref{fig:Transmission_Emission_LAr_Fresnel_Divergence_corrected} shows a spectrum of the transmission through 11.6\,cm pure liquid argon measured with light from a deuterium lamp. This spectrum has been obtained by a division of the data shown in figure\,\ref{fig:transmission_lar} by the curve in figure\,\ref{fig:fresnel_fokussierung_kombiniert}. On the current level of sensitivity no xenon-related effects could be observed and within the errors pure liquid argon is fully transparent down to the short wavelength cutoff of the system at 118\,nm. The error bars represent a linear sum of two uncertainties. Firstly, the statistical errors due to the intensity and the different exposure times of the light source (see error bars in figure\,\ref{fig:transmission_lar}). Secondly, the systematic errors due to the deviation from 1.0 in the test of the "fogging correction" (see figure\,\ref{fig:beschlag_benchmark} and section \ref{sec:fogging_correction} for details). 

Compared to figure\,6 in ref.\,\cite{Neumeier_epjc_2012} the results of the present measurements support the fact that the decreased transmission towards shorter wavelengths in ref.\,\cite{Neumeier_epjc_2012} was caused by a residual xenon impurity and not by liquid argon itself. If the reduced transmission towards shorter wavelengths would be an intrinsic feature of liquid argon itself this effect would be strongly increased in the present experiment since the length of the inner cell has been doubled compared to ref.\,\cite{Neumeier_epjc_2012}. This demonstrates that xenon was removed more efficiently due to the improved distillation procedure. We already measured the transmission of systematically xenon-doped liquid argon and the results are published in ref.\,\cite{neumeier_epl_vuv_transmission}. Only a wavelength-resolved measuring principle allows to identify impurities like xenon (or others) in nominally pure argon. Dependening on concentration and composition impurities affect both the emission as well as the transmission (see refs.\,\cite{neumeier_epl_vuv_ir,neumeier_epl_vuv_transmission}). Measuring the attenuation length in a wavelength-integrated way (e.g. ref.\,\cite{Ishida}) can not disentangle the influence of emission and absorption and therefore leads to unpredictable results. 

After all corrections the transmission above $\sim 160$\,nm is still $\sim 4$\,\% too high (see figure\,\ref{fig:Transmission_Emission_LAr_Fresnel_Divergence_corrected} (upper panel)). However, again it has to be emphasized that these corrections (Fresnel and "finite divergence effect") strongly rely on the refractive index of liquid argon which has not yet been measured. A density scaling of the refractive index from the gas phase to the liquid phase seems reasonable but for an explanation of the remaining $4$\,\% transmission above a value of 1.0 a precise measurement would have to be performed. Furthermore, it also has to be emphasized that below 140\,nm no measurement of the refractive index of gaseous argon has been carried out (see figure\,\ref{fig:Brechungsindex_GAr}). The scintillation wavelength of liquid argon peaks at $\sim127$\,nm (see figure\,\ref{fig:Transmission_Emission_LAr_Fresnel_Divergence_corrected}, lower panel). No measured information about the refractive index in the gas as well as in the liquid phase is available at these wavelengths. Using the refractive index of liquid argon obtained from an extrapolation by Grace and Nikkel \cite{Grace_Nikkel} (n=1.45 instead of n=1.34 from figure \ref{fig:Brechungsindex_LAr_MgF2} at a wavelength of 128 nm) to correct our raw data leads to a reduction of the transmission in figure \ref{fig:Transmission_Emission_LAr_Fresnel_Divergence_corrected} from 1.003 to 0.975 at 128 nm. This shift is too small to explain a Rayleigh scattering length of ($55\pm5$)\,cm \cite{Grace_Nikkel} (green curve in figure \ref{fig:Transmission_Emission_LAr_Fresnel_Divergence_corrected}) with our measured results. The red curve in figure\,\ref{fig:Transmission_Emission_LAr_Fresnel_Divergence_corrected} (upper panel) shows the expected transmission according to a Rayleigh scattering length of 90\,cm at 128\,nm \cite{Rayleigh_Seidel}. The measurement indicates a decreasing trend of the transmission towards the short wavelength end of the spectrum as may be the case due to the wavelength-dependent index of refraction or Rayleigh scattering. A scaling of the measurement to a transmission of 1.0 for wavelengths longer than 160\,nm indicates that the Rayleigh scattering length in liquid argon is longer than 90\,cm since the red line in figure\,\ref{fig:Transmission_Emission_LAr_Fresnel_Divergence_corrected} (upper panel) is below the lower limits of the error bars. However, a Rayleigh scattering length of 90\,cm can neither be confirmed nor excluded as long as no measured values for the wavelength-dependent refractive index of liquid argon are available. The black asterisk at 128\,nm in figure\,\ref{fig:Transmission_Emission_LAr_Fresnel_Divergence_corrected} shows the expected (in a 11.6\,cm long layer of liquid argon) transmission value at 128\,nm for an attenuation length of ($66\pm3$)\,cm measured by Ishida et al. \cite{Ishida}. However, a very conservative estimate based on the lower limits of the errors in the wavelength region from 122 to 135\,nm (scintillation region of liquid argon) in figure\,\ref{fig:Transmission_Emission_LAr_Fresnel_Divergence_corrected} shows a transmission which has to be better than $\sim0.9$ even if the data are scaled to match 1.0 for wavelengths longer than 160\,nm. Consequently, a lower limit for the attenuation length can be calculated to be $\sim1.10$\,m considering the length of 11.6\,cm for the inner cell.  

\begin{figure}
 \centering
 \includegraphics[width=\columnwidth]{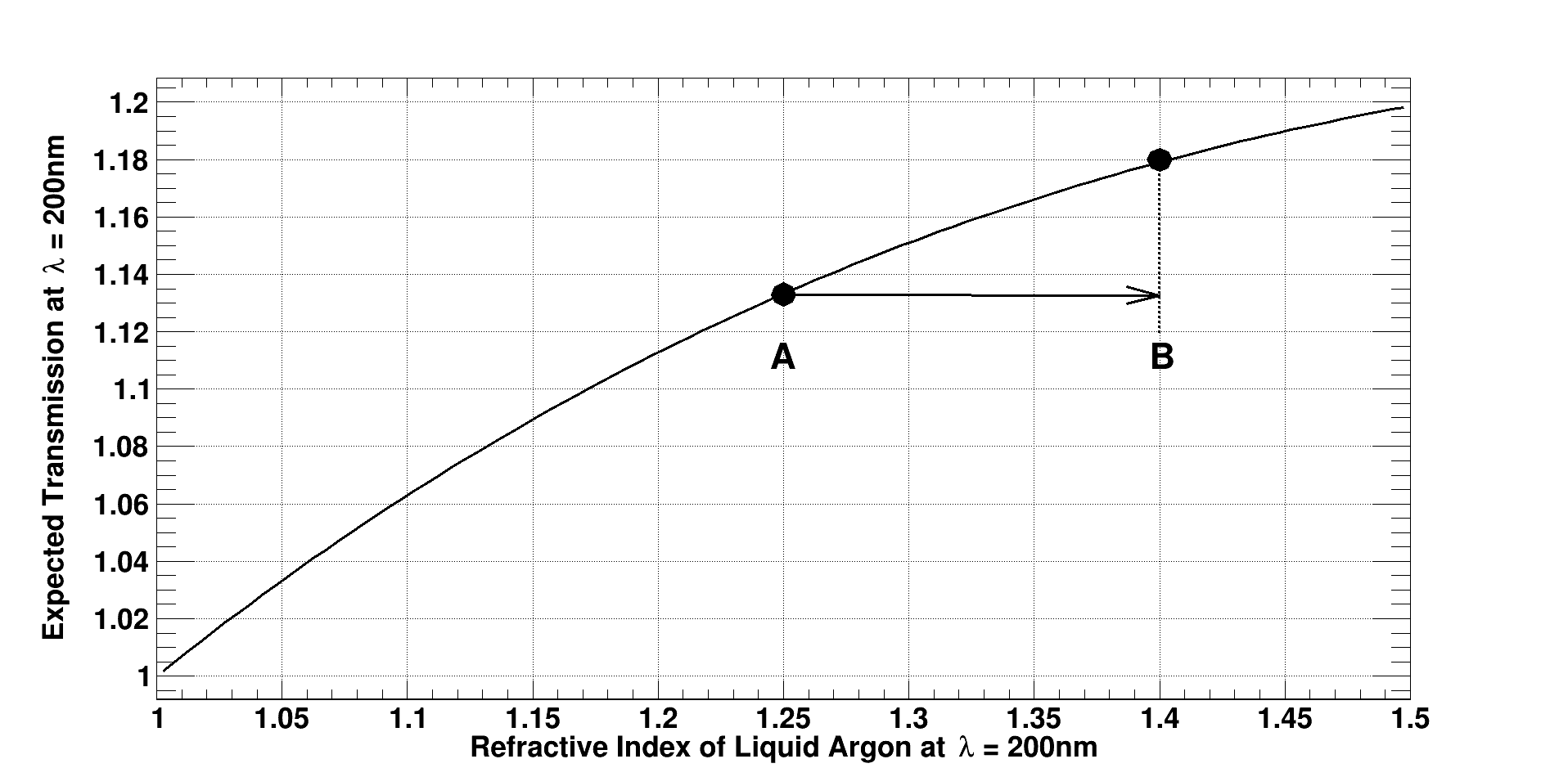}
 \caption{\textit{The expected transmission (according to "Fresnel and finite divergence effect") at a wavelength of 200\,nm versus the refractive index of liquid argon at 200\,nm is shown. For the calculation a refractive index of MgF$_{2}$ of 1.42 at a wavelength of 200\,nm has been used (see ref.\,\cite{laporte}). Point A denotes the refractive index of liquid argon at 200\,nm obtained from figure\,6. With a refractive index of liquid argon of 1.25 only an increased transmission of $\sim13$\,\% can be explained. If the measured increase of $\sim18$\,\% at a wavelength of 200\,nm (see figure\,4 and point B) has to be explained, the refractive index of liquid argon has to be changed from 1.25 to a value of 1.4 (black horizontal arrow). This corresponds to a relative change by 12\,\%.}}
 \label{fig:abschaetzung_brechungsindex}
\end{figure}

Another interesting question to be answered is the actual influence of the refractive index of liquid argon on the measured transmission. Figure\,\ref{fig:abschaetzung_brechungsindex} shows the expected transmission versus the refractive index of liquid argon at a wavelength of 200\,nm. We have chosen 200\,nm as an example since there is a measured value of the refractive index of the MgF$_{2}$ window available (see figure\,\ref{fig:Brechungsindex_LAr_MgF2} and ref.\,\cite{laporte}). The trend of the curve can be understood in the following way: If the refractive index of liquid argon would be 1.0 (i.e., the inner cell is evacuated) the transmission has to be 1.0. The curve has been calculated by a product of equations \ref{eq:fresnel_senkrecht} and \ref{eq:strahlfokussierung} (Fresnel and finite divergence) with a fixed refractive index of MgF$_{2}$ (1.42 at 200\,nm \cite{laporte}). The calculation has been performed with the parameters given by the experimental setup with the deuterium light source (x=17.5\,cm, y=\,11.6\,cm, z=36.5\,cm) for a variation of the refractive index of liquid argon from 1 to 1.5. Point A in figure\,\ref{fig:abschaetzung_brechungsindex} indicates the refractive index of liquid argon at 200\,nm from figure\,\ref{fig:Brechungsindex_LAr_MgF2} which has been used to obtain the corrected data in figure\,\ref{fig:Transmission_Emission_LAr_Fresnel_Divergence_corrected}. Point B in figure\,\ref{fig:abschaetzung_brechungsindex} indicates the measured transmission at 200\,nm from figure\,\ref{fig:transmission_lar}. At 200\,nm the raw data transmission has a value of 1.18. If the transmission above a value of 1.0 in figure\,\ref{fig:transmission_lar} at 200\,nm can be fully explained by the refractive index of liquid argon, the index of refraction has to be changed from $\sim 1.25$ to $\sim 1.4$ (from point A to B). This corresponds to a relative change of 12\,\%. 

An increase of the refractive index of liquid argon by about 12\,\% at a wavelength of 200\,nm could scale the data to 1.0. A similar argumentation can be given for the other wavelengths where the transmission is above 1.0. Since no measurements of the refractive index of liquid argon exist it can not be excluded that a wrong refractive index of liquid argon is the reason for the transmission still being $\sim4$\,\% too high at wavelengths longer than 160\,nm.

\section{Summary and Outlook}
The main goal of this work was to measure the wavelength-dependent attenuation length of liquid argon for its own scintillation light. In comparison to ref.\,\cite{Neumeier_epjc_2012} various unanswered questions could be addressed. The main result is that the decreased transmission towards shorter wavelengths in ref.\,\cite{Neumeier_epjc_2012} was caused by a residual xenon impurity and not by liquid argon itself. 

Transmission values above 1.0 in the raw data could partly be traced back to the Fresnel effect (see subsection 3.1) and a "finite divergence effect" of the light source used (see subsection 3.2). The "finite divergence effect" has been proven by transmission measurements extended to the UV, VIS and NIR regions. In addition, the beam profile has been measured using a He-Ne laser and a position-sensitive detector. The narrowing of the beam profile due to the "finite divergence effect" could directly be measured (see figure\,\ref{fig:ATIK_FWHM} and subsection 3.2).

The correction of the "fogging effect" (see section \ref{sec:fogging_correction} and ref.\,\cite{Neumeier_epjc_2012}) has been tested to be extremely reliable for wavelengths above 160\,nm. Below 160\,nm the correction introduces a systematic error on a $\sim5$\,\% level. The systematic errors of the "fogging correction" have been included in the data analysis.

Section \ref{sec:dataAnalysisMinorImprovement} shows that it is generally better to use light sources with a smooth emission spectrum or monochromatic light in transmission measurements.

The transmission of 11.6\,cm pure liquid argon is shown in the upper panel of figure\,\ref{fig:Transmission_Emission_LAr_Fresnel_Divergence_corrected} including all corrections. A very conservative estimate of the attenuation length based on the lower limits of the error bars in the region where liquid argon scintillates (see figure\,\ref{fig:Transmission_Emission_LAr_Fresnel_Divergence_corrected}, lower panel) leads to an attenuation length of more than $\sim1.10$\,m.

However, one major issue which could not be addressed is the lack of information on the wavelength-dependent refractive index of liquid argon. The influence of the refractive index on the transmission measurements in the vacuum ultraviolet presented here is estimated in figure\,\ref{fig:abschaetzung_brechungsindex} for a wavelength of 200\,nm. This leads to the conclusion that a variation of the refractive index of liquid argon on a 12\,\% level could scale the data to 1.0 for wavelengths longer than $\sim200$\,nm.

A major improvement of the experimental setup would be an increased length of the inner cell to about one meter to increase the sensitivity. A light source with a smooth emission spectrum and parallel light through the inner cell to eliminate the "finite divergence effect" would be superior compared to the equipment used in the present experiment. A vertical positioning of the inner cell with different filling heights of the liquid could eliminate all the issues which are related to the so far not measured refractive index of liquid argon.

However, based on the growing importance of liquid noble gases in particle detectors and especially due to the increasing detector volumes a wavelength-resolved measurement of the refractive index in the wavelength regions where liquid argon and xenon scintillate should also be performed.

\section*{Acknowledgements}
This research was supported by the DFG cluster of excellence "Origin and Structure of the Universe" (www.universe-cluster.de) and by the Maier-Leibnitz-Laboratorium in Garching.

\end{document}